\documentclass[11pt]{article}%
\usepackage{amssymb,amsmath,xcolor,graphicx,xspace,colortbl, rotating}
\usepackage{amsfonts}
\usepackage{amsthm}
\usepackage{caption}
\usepackage[mag=1000,portrait]{geometry}
\usepackage{indentfirst}
\usepackage[onehalfspacing]{setspace}
\usepackage{subcaption}
\usepackage{booktabs}
\usepackage{amsmath}
\usepackage{amssymb}
\usepackage{graphicx}%
\providecommand{\U}[1]{\protect\rule{.1in}{.1in}}
\DeclareGraphicsExtensions{.pdf,.eps,.ps,.png,.jpg,.jpeg}
\providecommand{\U}[1]{\protect\rule{.1in}{.1in}}
\newtheorem{theorem}{Theorem}

\newtheorem{assumption}{Assumption}
\newtheorem{corollary}{Corollary}

\newtheorem{lemma}{Lemma}

\DeclareMathOperator*{\argmax}{arg\,max}
\DeclareMathOperator*{\argmin}{arg\,min}
\begin{document}

\author{$%
\begin{array}
[c]{ccc}%
\text{Bruce E. Hansen}\thanks{Hansen thanks the National Science Foundation
and the Phipps Chair for research support.} & \hspace{0.2in} & \text{Seojeong
Lee}\thanks{Lee acknowledges that this research was supported under the
Australian Research Council Discovery Early Career Researcher Award (DECRA)
funding scheme (project number DE170100787).}\\
\text{University of Wisconsin} & \hspace{0.2in} & \text{University of New
South Wales}%
\end{array}
$ \medskip}
\title{\textbf{Asymptotic Theory for Clustered Samples}}
\date{February 2019$\footnote{We thank the Co-Editor Han Hong and two referees
for helpful comments on a previous version, and Morten Nielsen and James
MacKinnon for valuable conversations and suggestions.}$}
\maketitle

\begin{abstract}
We provide a complete asymptotic distribution theory for clustered data with a
large number of independent groups, generalizing the classic laws of large
numbers, uniform laws, central limit theory, and clustered covariance matrix
estimation. Our theory allows for clustered observations with heterogeneous
and unbounded cluster sizes. Our conditions cleanly nest the classical results
for i.n.i.d. observations, in the sense that our conditions specialize to the
classical conditions under independent sampling. We use this theory to develop
a full asymptotic distribution theory for estimation based on linear
least-squares, 2SLS, nonlinear MLE, and nonlinear GMM.

\end{abstract}

\pagebreak

\section{Introduction}

Clustered samples are widely used in current applied econometric practice.
Despite this dominance, there is little formal large-sample theory for
estimation and inference. This paper provides such a foundation. We develop a
complete, rigorous, and easily-interpretable asymptotic distribution theory
for the \textquotedblleft large number of clusters\textquotedblright%
\ framework. Our theory allows heterogeneous and growing cluster sizes, but
requires that the number of clusters $G$ grows with sample size $n$. Our core
theory provides a weak law of large numbers (WLLN), central limit theorem
(CLT), and consistent clustered variance estimation for clustered sample
means. We also provide uniform laws of large numbers and uniform consistent
clustered variance estimation appropriate for the distribution theory of
nonlinear econometric estimators.

We apply this core theory to develop large sample distribution theory for
standard econometric estimators: linear least-squares, 2SLS, MLE, and GMM. For
each, we provide conditions for consistent estimation, asymptotic normality,
consistent covariance matrix estimation, and asymptotic distributions for
t-ratios and Wald statistics. The theory provided in this paper is the first
formal theory for such econometric estimators allowing for clustered dependence.

Our assumptions are minimal, requiring only uniform integrability for the WLLN
and squared uniform integrability for the CLT and clustered covariance matrix
estimators, plus the requirement that individual clusters are asymptotically
negligible. Our results show that there are inherent trade-offs in the
conditions between the allowed degree of heterogeneity in cluster sizes and
the number of finite moments.\ These trade-offs are least restrictive for the
WLLN, are more restrictive for the CLT and consistent cluster covariance
matrix estimation, and are strongest for CLTs applied to clustered second
moments. These trade-offs do not arise in the independent sampling context.

We show that under clustering the convergence rate depends on the degree of
clustered dependence. Convergence rates may equal the square root of the
sample size, the square root of the number of clusters, be a rate in between
these two, or even slower than both. Our assumptions and theory allow for
these possibilities. This is in contrast to the existing literature, which
imposes specific rate assumptions. One useful finding is that the rate does
not need to be known by the user; the asymptotic distribution of t-ratios and
Wald statistics does not depend on the underlying rate of convergence. This
generalizes similar results in C. Hansen (2007) and related results in
Tabord-Meehan (2018).

This paper makes the following technical contributions. We show that the key
to extending the classical WLLN and CLT to cluster-level data is developing
uniform integrability bounds for cluster sums. To allow for arbitrary
within-cluster dependence, this means that such bounds will be scaled by
cluster sizes. This leads to bounds on the degree of cluster size
heterogeneity which can be allowed under cluster dependence. Some of the most
difficult technical work presented here is the extension of classical results
to clustered covariance matrix estimators. These are not sample averages, but
rather average across clusters of squared cluster sums. Handling such
estimators requires a new technical treatment.

Clustered dependence in econometrics dates to the work of Moulton (1986,
1990), Liang and Zeger (1986), and in particular Arellano (1987), who proposed
the popular cluster-robust covariance matrix estimator. The method was
popularized by the implementation in Stata by Rogers (1994) and the
widely-cited paper of Bertrand, Duflo and Mullainathan (2004). Surveys can be
found in Wooldridge (2003), Cameron and Miller (2011, 2015), MacKinnon (2012,
2016), and textbook treatments in Angrist and Pischke (2009) and Wooldridge (2010).

The \textquotedblleft large $G$\textquotedblright\ asymptotic theory develops
normal approximations under the assumption that $G\rightarrow\infty$. The
earliest treatment appears in White (1984). Wooldridge (2010) asserts a
distribution theory under the assumption that the cluster sizes are fixed. C.
Hansen (2007) provides two sets of asymptotic results, including both
$\sqrt{G}$ and $\sqrt{n}$ convergence rates under two distinct assumptions on
the rate of convergence of the estimation variance. His results are derived
under the assumption that all clusters are identical in size. Carter, Schnepel
and Steigerwald (2017) provided asymptotic results allowing for heterogeneous
clusters, but their results are limited by atypical regularity conditions.
Independently of this paper, Djogbenou, MacKinnon, and Nielsen (2018) have
provided a rigorous asymptotic theory for heterogeneous clusters, with similar
but stronger regularity conditions than ours. Their primary focus is theory
for regression wild bootstrap, while our focus is regularity conditions for
general econometric estimators.

An alternative to the \textquotedblleft large $G$\textquotedblright%
\ asymptotic is the \textquotedblleft fixed $G$\textquotedblright\ framework,
which leads to a non-normal inference theory. Contributions to this literature
include C. Hansen (2007), Bester, Conley and C. Hansen (2011), and Ibragimov
and M\"{u}eller (2010, 2016). A related paper is Conley and Taber (2011) which
provide an asymptotic theory under the assumption of a small number of groups
with policy changes. Canay, Romano, and Shaikh (2017) provide approximate
randomization tests.

Small sample approaches to cluster robust inference include Donald and Lang
(2007), Imbens and Koles\'{a}r (2016), and Young (2016). Bootstrap approaches
are provided by Cameron, Gelbach and Miller (2008), and MacKinnon and Webb
(2017, 2018).

A recent contribution which develops cluster-robust inference for GMM is Hwang (2017).

The organization of the paper is as follows. After Section 2, which introduces
cluster sampling, Sections 3-8 cover the core asymptotic theory, providing
rigorous conditions for the WLLN (Section 3), rates of convergence (Section
4), the CLT (Section 5), cluster-robust covariance matrix estimation (Section
6), the ULLN (Section 7), and the CLT for clustered second moments (Section
8). Following this, we provide the distribution theory for the core
econometric estimators, specifically linear regression and 2SLS (Section 9),
Maximum Likelihood (Section 10), and GMM\ (Section 11). Each of these latter
sections are written self-sufficiently, so they can be used directly by
readers. Proofs of the core theorems are provided in the Appendix, and proofs
for the applications are provided in the Supplemental Appendix.

\section{Cluster Sampling\label{cluster}}

The observations are $X_{i}\in%
\mathbb{R}
^{p}$ for $i=1,...,n$. They are grouped into $G$ mutually independent known
clusters, indexed $g=1,...,G$, where the $g^{th}$ cluster has $n_{g}$
observations. The clustering can be due to the sampling scheme, or done by the
researcher due to known correlation structures. The number of observations
$n_{g}$ per cluster (the \textquotedblleft cluster sizes\textquotedblright)
may vary across clusters. The total number of observations are $n=\sum
_{g=1}^{G}n_{g}$. It will also be convenient to double-index the observations
as $X_{gj}$ for $g=1,...,G$ and $j=1,...,n_{g}$.

As is conventional in the clustering literature, the only dependence
assumption we make is that the observations are independent across clusters,
while the dependence within each cluster is unrestricted. Furthermore, we do
not require that the observations or clusters come from identical
distributions. Thus our framework includes i.n.i.d (independent, not
necessarily identically distributed) as the special case $n_{g}=1$.

The notation and assumptions allow for linear panel data models with
cluster-specific fixed effects. In this case the observations $X_{gj}$ should
be viewed as clustered-demeaned observations. Another common application is
linear panel data models with both cluster-specific and time-specific fixed
effects. Our assumptions do not cover this case as removing the time effects
will induce cross-cluster correlations. This is essentially \textquotedblleft
multiway\textquotedblright\ clustering and requires different methods. See
MacKinnon, Nielsen and Webb (2017).

Our distributional framework is asymptotic as $n$ and $G$ simultaneously
diverge to infinity. This is typically referred to as the \textquotedblleft
large $G$\textquotedblright\ framework. Our assumptions, however, will allow
$G$ to diverge at a rate slower than $n$, by allowing the cluster sizes
$n_{g}$ to diverge. This is in contrast to the early asymptotic theory for
clustering, which implicitly assumed that the cluster sizes were bounded.

Our theory assumes that the clusters are known, and observations are
independent across clusters. This is a substantive restriction. Alternatively,
it may be possible to develop a distribution theory which allows weak
dependence across clusters, but we do not do so here.

A word on notation. For a vector $a$ let $\left\Vert a\right\Vert =\left(
a^{\prime}a\right)  ^{1/2}$ denote the Euclidean norm. For a positive
semi-definite matrix~$A$ let $\lambda_{\min}(A)$ and $\lambda_{\max}(A)$
denote its smallest and largest eigenvalue, respectively. For a general matrix
$A$ let $\left\Vert A\right\Vert =\sqrt{\lambda_{\max}\left(  A^{\prime
}A\right)  }$ denote the spectral norm. For a positive semi-definite
matrix~$A$ let $A^{1/2}$ denote the symmetric square root matrix such that
$A^{1/2}A^{1/2}=A$. We let $C$ denote a generic positive constant, that may be
different in different uses.

\section{Weak Law of Large Numbers}

For our core theory (WLLN \& CLT), we focus on the sample mean $\overline
{X}_{n}=\frac{1}{n}\sum_{i=1}^{n}X_{i}$ as an estimator of $E\overline{X}_{n}%
$. It will be convenient to define the cluster sums%
\[
\widetilde{X}_{g}=\sum_{j=1}^{n_{g}}X_{gj}%
\]
which are mutually independent under clustered sampling. The sample mean can
then be written as
\[
\overline{X}_{n}=\frac{1}{n}\sum_{g=1}^{G}\widetilde{X}_{g}.
\]

We use the following regularity condition.

\begin{assumption}
\label{A1} As $n\rightarrow\infty$%
\begin{equation}
\max\limits_{g\leq G}\dfrac{n_{g}}{n}\rightarrow0.
\end{equation}

\end{assumption}

\begin{theorem}
\label{wlln}(WLLN for clustered means). If Assumption \ref{A1} holds and
\begin{equation}
\lim_{M\rightarrow\infty}\sup_{i}\left(  E\left\Vert X_{i}\right\Vert 1\left(
\left\Vert X_{i}\right\Vert >M\right)  \right)  =0 \label{ui1}%
\end{equation}
then as $n\rightarrow\infty$,
\begin{equation}
\left\Vert \overline{X}_{n}-E\overline{X}_{n}\right\Vert \xrightarrow{p}0.
\label{wlln1}%
\end{equation}

\end{theorem}

The condition (\ref{ui1}) states that $X_{i}$ is uniformly
integrable\footnote{A referee points out that the sup in (\ref{ui1}) could be
weakened to an average. However our later results will use uniform
integrability conditions similar to (\ref{ui1}) so we state all results in
this format.}. This condition is identical to the standard condition for the
WLLN for independent heterogeneous observations, and thus Theorem \ref{wlln}
is a direct generalization of the WLLN for i.n.i.d. samples. (\ref{ui1})
simplifies to $E\left\Vert X_{i}\right\Vert <\infty$ when the observations
have identical marginal distributions. A sufficient condition allowing for
distributional heterogeneity is $\sup_{i}E\left\Vert X_{i}\right\Vert
^{r}<\infty$ for some $r>1$.

Assumption \ref{A1} states that each cluster size $n_{g}$ is asymptotically
negligible. This implies $G\rightarrow\infty$, so we do not explicitly need to
list the latter as an assumption. Assumption \ref{A1} allows for considerable
heterogeneity in cluster sizes. It allows the cluster sizes to grow with
sample size, so long as the growth is not proportional. For example, it allows
clusters to grow at the rate $n_{g}=n^{\alpha}$ for $0\leq\alpha<1$.

Assumption \ref{A1} is necessary for parameter estimation consistency while
allowing arbitrary within-cluster dependence. Otherwise a single cluster could
dominate the sample average. To see this, suppose that there is a cluster
$\ell$ such that all observations within the cluster are identical, so that
$X_{\ell j}=Z_{\ell}$ for some non-degenerate random variable $Z_{\ell}$, and
that this cluster violates Assumption \ref{A1}, so that $n_{\ell}/n\rightarrow
c>0$. Suppose for all other clusters that $EX_{gj}=0$ and $n_{g}%
/n\rightarrow0$. Then $\overline{X}_{n}\xrightarrow{p}Z_{\ell}$ and is
inconsistent. Thus Assumption \ref{A1} is necessary for the WLLN (\ref{wlln1})
if we allow for unstructured cluster heterogeneity.

Assumption \ref{A1} is equivalent to the condition%
\begin{equation}
\frac{\sum_{g=1}^{G}n_{g}^{2}}{n^{2}}\rightarrow0. \label{n2n}%
\end{equation}
To see this, first observe that since $\sum_{g=1}^{G}n_{g}=n$, the
left-hand-side of (\ref{n2n}) is smaller than $\max\limits_{g\leq G}%
n_{g}/n\rightarrow0$ under Assumption \ref{A1}. Thus Assumption \ref{A1}
implies (\ref{n2n}). Second,
\[
\max\limits_{g\leq G}\dfrac{n_{g}}{n}=\left(  \max\limits_{g\leq G}%
\dfrac{n_{g}^{2}}{n^{2}}\right)  ^{1/2}\leq\left(  \sum_{g=1}^{G}\dfrac
{n_{g}^{2}}{n^{2}}\right)  ^{1/2}\rightarrow0
\]
under (\ref{n2n}). Thus (\ref{n2n}) implies Assumption \ref{A1}, so the two
are equivalent.

\section{Rate of Convergence\label{rates}}

Under i.i.d. sampling the rate of convergence of the sample mean is $n^{-1/2}%
$. Clustering can alter the rate of convergence. In this section we explore
possible rates of convergence. From the work of C. Hansen (2007) it has been
understood that if the dependence within each cluster is weak then the rate of
convergence would be the i.i.d. rate $n^{-1/2}$ but if the dependence within
each cluster is strong then the rate of convergence would be determined by the
number of clusters: $G^{-1/2}$. What we now show is that the rate of
convergence can be in between or even slower than these rates.

The convergence rate can be calculated as the standard deviation of the sample
mean. For simplicity we focus on the scalar case $p=1$. The standard deviation
of $\overline{X}_{n}$ is
\[
\text{sd}\left(  \overline{X}_{n}\right)  =\frac{1}{n}\left(  \sum_{g=1}%
^{G}\text{var}(\widetilde{X}_{g})\right)  ^{1/2}.
\]

We now consider several examples. For our first four we take the case where
the clusters are all the same size: $n_{g}=n^{\alpha}$ for $0<\alpha<1$. In
this case the number of clusters is $G=n^{1-\alpha}$.

We first consider a case where the convergence is the i.i.d. rate $n^{-1/2}$.

\bigskip

\noindent\textbf{Example 1}. The observations are independent within each
cluster and $\text{var}(X_{i})=1$. Then
\[
\text{var}(\widetilde{X}_{g})=n_{g}=n^{\alpha}%
\]
and
\[
\text{sd}\left(  \overline{X}_{n}\right)  =n^{-1/2}.
\]

\bigskip

The $n^{-1/2}$ rate extends to any case where the within-cluster dependence is
weak, including autoregressive and moving average dependence.

Our second example is a case where the convergence is determined by the number
of clusters.

\bigskip

\noindent\textbf{Example 2}. The observations are identical within each
cluster (e.g. perfectly correlated) and $\text{var}(X_{i})=1$. Then
\[
\text{var}(\widetilde{X}_{g})=n_{g}^{2}=n^{2\alpha}%
\]
and
\[
\text{sd}\left(  \overline{X}_{n}\right)  =n^{-(1-\alpha)/2}=G^{-1/2}.
\]

\bigskip

The assumption that the observations are perfectly correlated is not essential
to obtain the $G^{-1/2}$ rate. What is important is that there is a common
component to the observations within a cluster.

Our third example is a case where the convergence rate is in between the above
two cases. Not surprisingly, it can obtained by constructing strong but
decaying within-cluster dependence.

\bigskip

\noindent\textbf{Example 3}. The observations are correlated within each
cluster with $\text{var}(X_{i})=1$ and $\text{cov}(X_{gj},X_{gl})=1/|j-l|$.
Then%
\[
\text{var}(\widetilde{X}_{g})\sim n_{g}\log n_{g}\sim n^{\alpha}\log n
\]
and
\[
\text{sd}\left(  \overline{X}_{n}\right)  \sim\sqrt{\log n/n}.
\]
Furthermore, $G\text{var}\left(  \overline{X}_{n}\right)  \rightarrow0.$ Thus
$\text{sd}\left(  \overline{X}_{n}\right)  $ converges at a rate in between
$n^{-1/2}$ and $G^{-1/2}$.

\bigskip

Our next two examples are somewhat surprising. They are cases where the
convergence rate is slower than both $n^{-1/2}$ and $G^{-1/2}$.

\bigskip

\noindent\textbf{Example 4}. The observations follow random walks within each
cluster: $X_{gj}=X_{gj-1}+\varepsilon_{gj}$ with $\varepsilon_{gj}$ i.i.d.
$(0,1)$ and $X_{g0}=0.$ Then
\[
\text{var}(\widetilde{X}_{g})\sim n_{g}^{3}%
\]
and
\[
\text{sd}\left(  \overline{X}_{n}\right)  \sim n^{\alpha-1/2}.
\]
Thus $\text{sd}\left(  \overline{X}_{n}\right)  $ converges at a rate slower
than both $n^{-1/2}$ and $G^{-1/2}$.

\bigskip

\noindent\textbf{Example 5}. The clusters are of two sizes, $n_{g}=1$ and
$n_{g}=n^{\alpha}$. There are $G_{1}=n/2$ of the first type and $G_{2}%
=n^{1-\alpha}/2$ of the second type. (So $G=G_{1}+G_{2}=O\left(  n\right)  $.)
Within each cluster the observations are identical and have unit variances.
var$(\widetilde{X}_{g})$ for the two types of clusters are $1$ and
$n^{2\alpha}$, respectively. Then
\[
\text{sd}\left(  \overline{X}_{n}\right)  =\left(  \frac{G_{1}+G_{2}%
n^{2\alpha}}{n^{2}}\right)  ^{1/2}=\left(  \frac{1+n^{\alpha}}{2n}\right)
^{1/2}=O\left(  n^{-(1-\alpha)/2}\right)  .
\]
Thus $\text{sd}\left(  \overline{X}_{n}\right)  $ converges at at a rate
slower than both $n^{-1/2}$ and $G^{-1/2}$.

\bigskip

The final example illustrates the importance of considering heterogeneous
cluster sizes. The reason why the convergence rate is slower than both
$n^{-1/2}$ and $G^{-1/2}$ is because the number of clusters is determined by
the large number of small clusters, but the convergence rate is determined by
the (relatively) small number of large clusters.

What we have seen is that the convergence rate $\text{sd}\left(  \overline
{X}_{n}\right)  $ can equal the square root of sample size $n^{-1/2}$, can
equal the square root of the number of groups $G^{-1/2}$, can be in between
$G^{-1/2}$ and $n^{-1/2}$, or can be slower than both $n^{-1/2}$ and
$G^{-1/2}$.

When $\overline{X}_{n}$ is a vector, it is likely that its elements converge
at different rates since they can have different within-cluster correlation
structures. For example, some variables could be independent within clusters
while others could be identical within clusters.

These examples show that under cluster dependence the convergence rate is
context-dependent and variable-dependent, and it is therefore important to
allow for general rates of convergence and to not impose arbitrary rates in
asymptotic analysis.

\section{Central Limit Theory}

\label{sCLT}

Under i.i.d. sampling the standard deviation of the sample mean is of order
$O(n^{-1/2})$, so $\sqrt{n}$ is the appropriate scaling to obtain the central
limit theorem (CLT). As discussed in the previous section, clustering can
alter the rate of convergence, so it is essential to standardize the sample
mean by the actual variance rather than an assumed rate. The variance matrix
of $\sqrt{n}\overline{X}_{n}$ is%
\begin{align*}
\Omega_{n}  &  =E\left(  n\left(  \overline{X}_{n}-E\overline{X}_{n}\right)
\left(  \overline{X}_{n}-E\overline{X}_{n}\right)  ^{\prime}\right) \\
&  =\frac{1}{n}\sum_{g=1}^{G}E\left(  \left(  \widetilde{X}_{g}-E\widetilde{X}%
_{g}\right)  \left(  \widetilde{X}_{g}-E\widetilde{X}_{g}\right)  ^{\prime
}\right)  .
\end{align*}

We use the following regularity condition.

\begin{assumption}
\label{A2} For some $2\leq r<\infty$
\begin{equation}
\dfrac{\left(  \sum_{g=1}^{G}n_{g}^{r}\right)  ^{2/r}}{n}\leq C<\infty,
\label{nbound}%
\end{equation}%
\begin{equation}
\max\limits_{g\leq G}\dfrac{n_{g}^{2}}{n}\rightarrow0, \label{ng2n}%
\end{equation}
as $n\rightarrow\infty.$
\end{assumption}

\begin{theorem}
\label{clt}(CLT) If for some $2\leq r<\infty$ Assumption \ref{A2} holds,
\begin{equation}
\lim_{M\rightarrow\infty}\sup_{i}\left(  E\left\Vert X_{i}\right\Vert
^{r}1\left(  \left\Vert X_{i}\right\Vert >M\right)  \right)  =0,\label{uir}%
\end{equation}
and%
\begin{equation}
\lambda_{n}=\lambda_{\min}\left(  \Omega_{n}\right)  \geq\lambda
>0,\label{lambdan}%
\end{equation}
then as $n\rightarrow\infty$%
\begin{equation}
\Omega_{n}^{-1/2}\sqrt{n}\left(  \overline{X}_{n}-E\overline{X}_{n}\right)
\xrightarrow {d}N\left(  \mathbf{0},I_{p}\right)  .\label{clt1}%
\end{equation}

\end{theorem}

Theorem \ref{clt} provides a CLT for cluster samples which generalizes the
classic CLT for independent heterogeneous samples. The latter holds with $r=2
$, $n_{g}=1$ and $G=n$.

Assumption \ref{A2} and \eqref{uir} are stronger than Assumption \ref{A1} and
\eqref{ui1}, and thus the conditions for the CLT imply those for the WLLN.

The condition (\ref{uir}) states that $\left\Vert X_{i}\right\Vert ^{r}$ is
uniformly integrable. When $r=2$ this is similar to the Lindeberg condition
for the CLT under independent heterogeneous sampling. (\ref{uir}) simplifies
to $E\left\Vert X_{i}\right\Vert ^{r}<\infty$ when the observations have
identical marginal distributions. A sufficient condition allowing for
distributional heterogeneity is $\sup_{i}E\left\Vert X_{i}\right\Vert
^{s}<\infty$ for some $s>r\geq2$.

Assumption \ref{A2} (\ref{nbound}) is a restriction on the cluster sizes. It
involves a trade-off with the number of moments $r$. It is least restrictive
for large $r$, and more restrictive for small $r$. As $r\rightarrow\infty$ it
approaches $\max_{g\leq G}n_{g}^{2}/n = O(1)$, which is implied by Assumption
\ref{A2} (\ref{ng2n}).

Assumption \ref{A2} allows for growing and heterogeneous cluster sizes. For
example, it allows clusters to grow uniformly at the rate $n_{g}=n^{\alpha}$
for $0\leq\alpha\leq(r-2)/2(r-1)$. (Note that this requires the cluster sizes
to be bounded if $r=2$.) It also allows for only a small number of clusters to
grow. For example, suppose that $n_{g}=\overline{n}$ (bounded) for $G-K$
clusters and $n_{g}=G^{\alpha/2 }$ for $K$ clusters, with $K$ fixed. Then
Assumption \ref{A2} holds for any $\alpha<1$ and $r\geq2$.

Assumption \ref{A2} (\ref{nbound}) is implied by
\begin{equation}
\max\limits_{g\leq G}\dfrac{n_{g}}{n^{(r-2)/2(r-1)}}\leq C \label{nbound1}%
\end{equation}
and they are equivalent when the cluster sizes are homogeneous. In general,
however, (\ref{nbound}) is less restrictive than (\ref{nbound1}). For example,
when $r=2$, (\ref{nbound1}) requires the cluster sizes to be bounded, while
(\ref{nbound}) does not. (Consider the heterogeneous example given in the
previous paragraph. This satisfies (\ref{nbound}) but not (\ref{nbound1}) when
$r=2$.)

The condition (\ref{lambdan}) specifies that $\text{var}\left(  \sqrt{n}%
\alpha^{\prime}\overline{X}_{n}\right)  $ does not vanish for any conformable
vector $\alpha\neq0$. This excludes degenerate cases and perfect negative
within-cluster correlation. In general, if $X_{i}$ is non-degenerate then
(\ref{lambdan}) is not restrictive as there is no reasonable setting where it
will be violated. If $\overline{X}_{n}$ converges at rate $n^{-1/2}$ then
$\lambda_{n}=O(1)$ but when $\overline{X}_{n}$ converges at rate slower than
$n^{-1/2}$ then $\lambda_{n}$ will actually diverge with $n$. It should also
be mentioned that condition (\ref{lambdan}) allows the components of
$\Omega_{n}$ to converge at different rates.

Our proof of Theorem \ref{clt} actually uses the conditions%
\begin{equation}
\dfrac{\left(  \sum_{g=1}^{G}n_{g}^{r}\right)  ^{2/r}}{n\lambda_{n}}\leq
C<\infty\label{n1}%
\end{equation}
and%
\begin{equation}
\max\limits_{g\leq G}\dfrac{n_{g}^{2}}{n\lambda_{n}}\rightarrow0 \label{n2}%
\end{equation}
instead of (\ref{nbound})-(\ref{lambdan}). (\ref{n1})-(\ref{n2}) is weaker
than (\ref{nbound})-(\ref{lambdan}) when $\lambda_{n}$ diverges to infinity
(which occurs when $\overline{X}_{n}$ converges at a rate slower than
$n^{-1/2}$). Since the sequence $\lambda_{n}$ is unknown in an application it
is difficult to interpret the assumptions (\ref{n1})-(\ref{n2}). Hence we
prefer the assumptions (\ref{nbound})-(\ref{lambdan}).

The conditions (\ref{n1})-(\ref{n2}) may be stronger than necessary when
within-cluster dependence is weak, but are necessary under strong
within-cluster dependence. To see this, suppose that all observations within a
cluster are identical, so that $X_{gj}=Z_{g}$ and $Z_{g}$ has a finite
variance but no higher moments. Then the Lindeberg condition for the CLT can
be simplified to%
\[
\sum_{g=1}^{G}\frac{n_{g}^{2}}{n\lambda_{n}}E\left(  \left\Vert Z_{g}%
\right\Vert ^{2}1\left(  \left\Vert Z_{g}\right\Vert ^{2}\geq\frac
{n\lambda_{n}\varepsilon}{n_{g}^{2}}\right)  \right)  \rightarrow0
\]
for all $\varepsilon>0$. Each term in the sum must limit to zero, which
requires (\ref{n1})-(\ref{n2}) with $r=2$.

We now compare our conditions with those of Djogbenou, MacKinnon, and Nielsen
(2018). Their Assumption 3 states (in our notation) for $r\geq4$%
\begin{equation}
\max\limits_{g\leq G}\dfrac{n_{g}}{n^{(r-2)/2(r-1)}\lambda_{n}^{r/2(r-1)}%
}=o(1).\label{lambda2}%
\end{equation}
Equation (\ref{lambda2}) implies and is stronger than (\ref{n1}). Calculations
similar to those in our appendix show that $\lambda_{n}\leq O\left(  \max
_{g}n_{g}\right)  =O(n)$. So (\ref{lambda2}) also implies%
\[
\left(  \max\limits_{g\leq G}\dfrac{n_{g}^{2}}{n\lambda_{n}}\right)
^{1/2}=\max\limits_{g\leq G}\dfrac{n_{g}}{n^{(r-2)/2(r-1)}\lambda
_{n}^{r/2(r-1)}}\left(  \frac{\lambda_{n}}{n}\right)  ^{1/2(r-1)}=o\left(
1\right)
\]
which is (\ref{n2}). Thus our conditions (\ref{n1})-(\ref{n2}) are less
restrictive than their condition (\ref{lambda2}), and do not require $r\geq4$.

\section{Cluster-Robust Variance Matrix Estimation}

We now discuss cluster-robust covariance matrix estimation.

We first consider the case where $X_{i}$ is mean zero (or equivalently that
the mean is known). In this case the covariance matrix equals%
\[
\Omega_{n}=\frac{1}{n}\sum_{g=1}^{G}E\left(  \widetilde{X}_{g}\widetilde{X}%
_{g}^{\prime}\right)  .
\]
In this case a natural estimator is%
\[
\widetilde{\Omega}_{n}=\frac{1}{n}\sum_{g=1}^{G}\widetilde{X}_{g}%
\widetilde{X}_{g}^{\prime}.
\]

\begin{theorem}
\label{cov}Under the assumptions of Theorem \ref{clt}, if in addition
$EX_{i}=0$ then as $n\rightarrow\infty$%
\begin{equation}
\Omega_{n}^{-1/2}\widetilde{\Omega}_{n}\Omega_{n}^{-1/2}\xrightarrow{p}I_{p}
\label{v2}%
\end{equation}
and%
\begin{equation}
\widetilde{\Omega}_{n}^{-1/2}\sqrt{n}\overline{X}_{n}\xrightarrow {d}N\left(
\mathbf{0},I_{p}\right)  . \label{v3}%
\end{equation}

\end{theorem}

Theorem \ref{cov} shows that the cluster-robust covariance matrix estimator is
consistent, and replacing the covariance matrix in the CLT\ with the estimated
covariance matrix does not affect the asymptotic distribution. Implications of
(\ref{v3}) are that cluster-robust t-ratios are asymptotically standard
normal, and that cluster-robust Wald statistics are asymptotically chi-square
distributed with $p$ degrees of freedom.

Construction of practical covariance matrix estimators is context-specific,
depending on the mean structure. For example, suppose that $\mu=EX_{i}$ does
not vary across observations. In this case we can write
\[
\Omega_{n}=\frac{1}{n}\sum_{g=1}^{G}E\left(  \widetilde{X}_{g}\widetilde{X}%
_{g}^{\prime}\right)  -\frac{1}{n}\sum_{g=1}^{G}n_{g}^{2}\mu\mu^{\prime}.
\]
The natural estimator for $\mu$ is $\overline{X}_{n}$ and that for $\Omega
_{n}$ is%
\[
\widehat{\Omega}_{n}=\frac{1}{n}\sum_{g=1}^{G}\widetilde{X}_{g}\widetilde{X}%
_{g}^{\prime}-\frac{1}{n}\sum_{g=1}^{G}n_{g}^{2}\overline{X}_{n}\overline
{X}_{n}^{\prime}.
\]

\begin{theorem}
\label{cov2}Under the assumptions of Theorem \ref{clt}, if in addition
$\mu=EX_{i}$ does not vary across observations, then as $n\rightarrow\infty$%
\begin{equation}
\Omega_{n}^{-1/2}\widehat{\Omega}_{n}\Omega_{n}^{-1/2}\xrightarrow{p}I_{p}
\label{v4}%
\end{equation}
and%
\begin{equation}
\widehat{\Omega}_{n}^{-1/2}\sqrt{n}\left(  \overline{X}_{n}-\mu\right)
\xrightarrow {d}N\left(  \mathbf{0},I_{p}\right)  . \label{v5}%
\end{equation}

\end{theorem}

\section{Uniform Laws of Large Numbers}

Now consider a uniform WLLN. Consider functions $f(x,\theta)\in\mathbb{R}^{k}$
indexed on $\theta\in\Theta$ where $\Theta$ is compact. Define the sample
mean
\[
\overline{f}_{n}(\theta)=\frac{1}{n}\sum_{i=1}^{n}f(X_{i},\theta).
\]

The following result is an application of Theorem 3 of Andrews (1992).

\begin{theorem}
\label{ulln}(ULLN for clustered means). Suppose that Assumption \ref{A1} holds
and for each $\theta\in\Theta$%
\begin{equation}
\lim_{M\rightarrow\infty}\sup_{i}\left(  E\left\Vert f(X_{i},\theta
)\right\Vert 1\left(  \left\Vert f(X_{i},\theta)\right\Vert >M\right)
\right)  =0. \label{f1}%
\end{equation}
Suppose as well that for each $\theta_{1},\theta_{2}\in\Theta$
\begin{equation}
\left\Vert f(x,\theta_{1})-f(x,\theta_{2})\right\Vert \leq A(x)h\left(
\left\Vert \theta_{1}-\theta_{2}\right\Vert \right)  \label{lip1}%
\end{equation}
where $h(u)\downarrow0$ as $u\downarrow0$ and $\sup_{i}EA(X_{i})\leq C$. Then
$E\overline{f}_{n}(\theta)$ is continuous in $\theta$ uniformly over
$\theta\in\Theta$ and $n\geq1$, and as $n\rightarrow\infty$%
\begin{equation}
\sup_{\theta\in\Theta}\left\Vert \overline{f}_{n}(\theta)-E\overline{f}%
_{n}(\theta)\right\Vert \xrightarrow{p}0. \label{supu}%
\end{equation}

\end{theorem}

We also consider a uniform law for the clustered variance. Set $\mu
(\theta)=Ef(X_{i},\theta)$ so that it does not vary across observations. The
variance of $\sqrt{n}\overline{f}_{n}(\theta)$ is
\begin{align*}
\Omega_{n}(\theta)  &  =E\left(  n\left(  \overline{f}_{n}(\theta
)-E\overline{f}_{n}(\theta)\right)  \left(  \overline{f}_{n}(\theta
)-E\overline{f}_{n}(\theta)\right)  ^{\prime}\right) \\
&  =\frac{1}{n}\sum_{g=1}^{G}E\widetilde{f}_{g}(\theta)\widetilde{f}%
_{g}(\theta)-\frac{1}{n}\sum_{g=1}^{G}n_{g}^{2}\mu(\theta)\mu(\theta)^{\prime}%
\end{align*}
where $\widetilde{f}_{g}(\theta)=\sum_{j=1}^{n_{g}}f(X_{gj},\theta)$ are the
cluster sums. An appropriate estimator for $\Omega_{n}(\theta)$ is
\[
\widehat{\Omega}_{n}(\theta)=\frac{1}{n}\sum_{g=1}^{G}\widetilde{f}_{g}%
(\theta)\widetilde{f}_{g}(\theta)-\frac{1}{n}\sum_{g=1}^{G}n_{g}^{2}%
\overline{f}_{n}(\theta)\overline{f}_{n}(\theta)^{\prime}.
\]
In practice, a simpler estimator
\[
\widetilde{\Omega}_{n}(\theta)=\frac{1}{n}\sum_{g=1}^{G}\widetilde{f}%
_{g}(\theta)\widetilde{f}_{g}(\theta)^{\prime}%
\]
is often used if $\mu(\theta_{0})=0$ for $\theta_{0}\in interior\left(
\Theta\right)  $ and $\widehat{\theta}\xrightarrow{p}\theta_{0}$ for some
estimator $\widehat{\theta}$.

The following result is an extension of Theorem \ref{ulln} to the case of
clustered variance estimators. It also relies on Theorem 3 of Andrews (1992).

\begin{theorem}
\label{ullnv}(ULLN for clustered variance). Suppose that Assumption \ref{clt}
holds with $r=2$, $\mu(\theta)=Ef(X_{i},\theta)$ does not vary across $i$, for
each $\theta\in\Theta$,
\begin{equation}
\lim_{M\rightarrow\infty}\sup_{i}\left(  E\left\Vert f(X_{i},\theta
)\right\Vert ^{2}1\left(  \left\Vert f(X_{i},\theta)\right\Vert >M\right)
\right)  =0, \label{fr}%
\end{equation}
and for each $\theta_{1},\theta_{2}\in\Theta$ \eqref{lip1} holds with
$\sup_{i}EA(X_{i})^{2}\leq C$. Then as $n\rightarrow\infty$
\begin{equation}
\sup_{\theta\in\Theta}\left\Vert \widehat{\Omega}_{n}(\theta)-\Omega
_{n}(\theta)\right\Vert \xrightarrow{p}0. \label{vu2}%
\end{equation}
If $\mu(\theta)=0$, then as $n\rightarrow\infty$
\begin{equation}
\sup_{\theta\in\Theta}\left\Vert \widetilde{\Omega}_{n}(\theta)-\Omega
_{n}(\theta)\right\Vert \xrightarrow{p}0. \label{vu3}%
\end{equation}

\end{theorem}

\section{Central Limit Theorem for Clustered Second Moments}

Although our primary focus is the sample mean, the core theory can be extended
to statistics which are not sample means. In this section, we focus on the
vectorized variance estimators
\[
\overline{f}_{G}=\frac{1}{n}\sum_{g=1}^{G}\widetilde{f}_{g}%
\]
where
\[
\widetilde{f}_{g}=\widetilde{X}_{g}\otimes\widetilde{X}_{g}%
\]
or
\[
\widetilde{f}_{g}=\left(  \widetilde{X}_{g}-n_{g}\overline{X}_{n}\right)
\otimes\left(  \widetilde{X}_{g}-n_{g}\overline{X}_{n}\right)  .
\]
The WLLN for $\overline{f}_{G}$ holds by Theorem \ref{cov} (\ref{v2}) and
Theorem \ref{cov2} (\ref{v4}), and the ULLN for $\overline{f}_{G}$ holds by
Theorem \ref{ullnv}. However, the CLT given in Theorem \ref{clt} cannot be
applied to $\overline{f}_{G}$ because $\overline{f}_{G}$ cannot be written as
the sample mean over $i$. We provide the CLT for $\overline{f}_{G}$ below.
This is useful to establish asymptotic distributions of estimators in a
non-standard setting. For example, the asymptotic distribution of the
generalized method of moments (GMM) estimators depends on the limiting
distribution of the weight matrix when the moment condition is misspecified
(Hall and Inoue, 2003; Lee, 2014; Hansen and Lee, 2018).

Similar to the sample mean, the convergence rate of $\overline{f}_{G}$ can
vary under cluster dependence. Consider $\widetilde{f}_{g} = \widetilde{X}%
_{g}\otimes\widetilde{X}_{g}$ and assume $p=1$ for simplicity. The standard
deviation of $\overline{f}_{G}$ is
\[
\text{sd}\left(  \overline{f}_{G}\right)  =\frac{1}{n}\left(  \sum_{g=1}%
^{G}\text{var}\left(  \widetilde{X}_{g}\widetilde{X}_{g}\right)  \right)
^{1/2} = \frac{1}{n}\left(  \sum_{g=1}^{G}\sum_{j=1}^{n_{g}}\sum_{l=1}^{n_{g}%
}\text{var}\left(  X_{gj}X_{gl}\right)  \right)  ^{1/2}.
\]
Under i.i.d. sampling $\text{sd}\left(  \overline{f}_{G}\right)  =O\left(
n^{-1/2}\right)  $. Under the Examples 1 and 2 in Section \ref{rates}, the
convergence rate is $G^{-1/2}$.

Define the variance matrix of $\sqrt{n}\overline{f}_{G}$ as
\begin{align*}
\Omega_{n}  &  =E\left(  n\left(  \overline{f}_{G}-E\overline{f}_{G}\right)
\left(  \overline{f}_{G}-E\overline{f}_{G}\right)  ^{\prime}\right) \\
&  =\frac{1}{n}\sum_{g=1}^{G}E\left(  \left(  \widetilde{f}_{g}-E\widetilde{f}%
_{g}\right)  \left(  \widetilde{f}_{g}-E\widetilde{f}_{g}\right)  ^{\prime
}\right)  .
\end{align*}

We use the following regularity condition.

\begin{assumption}
\label{A3}For some $2\leq r<\infty$
\begin{equation}
\dfrac{\left(  \sum_{g=1}^{G}n_{g}^{2r}\right)  ^{2/r}}{n}\leq C<\infty,
\label{nboundv}%
\end{equation}%
\begin{equation}
\max\limits_{g\leq G}\dfrac{n_{g}^{4}}{n}\rightarrow0, \label{ng4n}%
\end{equation}
as $n\rightarrow\infty$.
\end{assumption}

Note that Assumption \ref{A3} is a strengthening of Assumption \ref{A2}.

\begin{theorem}
\label{cltv}(CLT for clustered variance) For some $2\leq r<\infty$ Assumption
\ref{A3} holds,
\begin{equation}
\lim_{M\rightarrow\infty}\sup_{i}\left(  E\left\Vert X_{i}\right\Vert
^{2r}1\left(  \left\Vert X_{i}\right\Vert >M\right)  \right)  =0, \label{uirv}%
\end{equation}
and%
\begin{equation}
\lambda_{n}=\lambda_{\min}\left(  \Omega_{n}\right)  \geq\lambda>0
\end{equation}
then as $n\rightarrow\infty$
\begin{equation}
\Omega_{n}^{-1/2}\sqrt{n}\left(  \overline{f}_{G}-E\overline{f}_{G}\right)
\xrightarrow {d}N\left(  \mathbf{0},I_{q}\right)  \label{clt2}%
\end{equation}
where $q=p^{2}$.
\end{theorem}

Finally we provide a CLT combining the previous results. For $Y_{i}%
\in\mathbb{R}^{s}$, $i=1,...,n$, obtained by cluster sampling, let
$\widetilde{\psi}_{g}$ be the stacked vector
\[
\widetilde{\psi}_{g}=\left(
\begin{array}
[c]{c}%
\widetilde{Y}_{g}\\
\widetilde{X}_{g}\\
\widetilde{X}_{g}\otimes\widetilde{X}_{g}%
\end{array}
\right)
\]
or
\[
\widetilde{\psi}_{g}=\left(
\begin{array}
[c]{c}%
\widetilde{Y}_{g}\\
\widetilde{X}_{g}\\
\left(  \widetilde{X}_{g}-n_{g}\overline{X}_{n}\right)  \otimes\left(
\widetilde{X}_{g}-n_{g}\overline{X}_{n}\right)
\end{array}
\right)
\]
and $\overline{\psi}_{G}=n^{-1}\sum_{g=1}^{G}\widetilde{\psi}_{g}$. Let the
variance matrix of $\sqrt{n}\overline{\psi}_{G}$ be
\[
\Omega_{n}=E\left(  n\left(  \overline{\psi}_{G}-E\overline{\psi}_{G}\right)
\left(  \overline{\psi}_{G}-E\overline{\psi}_{G}\right)  ^{\prime}\right)  .
\]
The following Corollary provides the CLT for the joint process. Since it
immediately follows from Theorems \ref{clt} and \ref{cltv}, the proof is omitted.

\begin{corollary}
\label{cltj} If for some $2\leq r<\infty$ Assumption \ref{A3} holds,
\begin{align*}
&  \lim_{M\rightarrow\infty}\sup_{i}\left(  E\left\Vert Y_{i}\right\Vert
^{r}1\left(  \left\Vert Y_{i}\right\Vert >M\right)  \right)  =0,\\
&  \lim_{M\rightarrow\infty}\sup_{i}\left(  E\left\Vert X_{i}\right\Vert
^{2r}1\left(  \left\Vert X_{i}\right\Vert >M\right)  \right)  =0,
\end{align*}
and
\[
\lambda_{\min}\left(  \Omega_{n}\right)  \geq\lambda>0,
\]
then as $n\rightarrow\infty$
\[
\Omega_{n}^{-1/2}\sqrt{n}\left(  \overline{\psi}_{G}-E\overline{\psi}%
_{G}\right)  \xrightarrow {d}N\left(  \mathbf{0},I_{q}\right)
\]
where $q=s+p+p^{2}$.
\end{corollary}

\section{Linear Regression and Two-Stage Least Squares}

It is useful to use cluster-level notation. Let $\boldsymbol{y}_{g}%
=(y_{g1},...,y_{gn_{g}})^{\prime}$, $\boldsymbol{X}_{g}=(\boldsymbol{x}%
_{g1},...,\boldsymbol{x}_{gn_{g}})^{\prime}$ and $\boldsymbol{Z}%
_{g}=(\boldsymbol{z}_{g1},...,\boldsymbol{z}_{gn_{g}})^{\prime}$ denote an
$n_{g}\times1$ vector of dependent variables, $n_{g}\times k$ matrix of
regressors, and $n_{g}\times l$ matrix of instruments for the $g^{th}$
cluster. A linear model can be written using cluster notation as%
\begin{align}
\boldsymbol{y}_{g}  &  =\boldsymbol{X}_{g}\boldsymbol{\beta}+\boldsymbol{e}%
_{g},\label{me}\\
\boldsymbol{X}_{g}  &  =\boldsymbol{Z}_{g}\boldsymbol{\gamma}+\boldsymbol{u}%
_{g},\label{fs}\\
E\left(  \boldsymbol{Z}_{g}^{\prime}\boldsymbol{e}_{g}\right)   &  =0\nonumber
\end{align}
where $\boldsymbol{e}_{g}$ is a $n_{g}\times1$ error vector. The case of
linear regression holds as the special case where $\boldsymbol{Z}%
_{g}=\boldsymbol{X}_{g}$ and $l=k$ (so that \eqref{fs} becomes identity).
Assume $l\geq k$. (\ref{me}) is the structural equation and (\ref{fs}) is the
first-stage equation.

The two-stage least squares (2SLS) estimator for $\boldsymbol{\beta}$ can be
written as
\[
\widehat{\boldsymbol{\beta}}=\left(  \sum_{g=1}^{G}\boldsymbol{X}_{g}^{\prime
}\boldsymbol{Z}_{g}\left(  \sum_{g=1}^{G}\boldsymbol{Z}_{g}^{\prime
}\boldsymbol{Z}_{g}\right)  ^{-1}\sum_{g=1}^{G}\boldsymbol{Z}_{g}^{\prime
}\boldsymbol{X}_{g}\right)  ^{-1}\left(  \sum_{g=1}^{G}\boldsymbol{X}%
_{g}^{\prime}\boldsymbol{Z}_{g}\left(  \sum_{g=1}^{G}\boldsymbol{Z}%
_{g}^{\prime}\boldsymbol{Z}_{g}\right)  ^{-1}\sum_{g=1}^{G}\boldsymbol{Z}%
_{g}^{\prime}\boldsymbol{y}_{g}\right)  .
\]
We first show consistency of $\widehat{\boldsymbol{\beta}}$. Define
\begin{align*}
Q_{n}  &  =\frac{1}{n}\sum_{g=1}^{G}E\left(  \boldsymbol{Z}_{g}^{\prime
}\boldsymbol{X}_{g}\right)  ,\\
W_{n}  &  =\frac{1}{n}\sum_{g=1}^{G}E\left(  \boldsymbol{Z}_{g}^{\prime
}\boldsymbol{Z}_{g}\right)  .
\end{align*}

\begin{theorem}
\label{2SLSc} If Assumption \ref{A1} holds, $Q_{n}$ has full rank $k$,
$\lambda_{\min}(W_{n})\geq C>0$, and either

\begin{enumerate}
\item $(y_{i},\boldsymbol{x}_{i},\boldsymbol{z}_{i})$ have identical marginal
distributions with finite second moments;

or

\item For some $r>2$, $\sup_{i}E\left|  y_{i}\right|  ^{r}<\infty$, $\sup
_{i}E\left\Vert \boldsymbol{x}_{i}\right\Vert ^{r}<\infty$, and $\sup
_{i}E\left\Vert \boldsymbol{z}_{i}\right\Vert ^{r}<\infty;$
\end{enumerate}

then as $n\rightarrow\infty$, $\widehat{\boldsymbol{\beta}}%
\xrightarrow{p}\boldsymbol{\beta}.$
\end{theorem}

Next we provide the asymptotic distribution. Define
\begin{align*}
\Omega_{n}  &  = \frac{1}{n}\sum_{g=1}^{G}E\left(  \boldsymbol{Z}_{g}^{\prime
}\boldsymbol{e}_{g}\boldsymbol{e}_{g}^{\prime}\boldsymbol{Z}_{g}\right)  ,\\
V_{n}  &  = \left(  Q_{n}^{\prime}W_{n}^{-1}Q_{n}\right)  ^{-1}Q_{n}^{\prime
}W_{n}^{-1}\Omega_{n}W_{n}^{-1}Q_{n}\left(  Q_{n}^{\prime}W_{n}^{-1}%
Q_{n}\right)  ^{-1}.
\end{align*}

The residuals for the $g^{th}$ cluster are%
\[
\widehat{\boldsymbol{e}}_{g}=\boldsymbol{y}_{g}-\boldsymbol{X}_{g}\widehat{
\boldsymbol{\beta}}.
\]

Define
\begin{align*}
\widehat{\Omega}_{n}  &  =\frac{1}{n}\sum_{g=1}^{G}\boldsymbol{Z}_{g}^{\prime
}\widehat{\boldsymbol{e}}_{g}\widehat{\boldsymbol{e}}_{g}^{\prime
}\boldsymbol{Z}_{g},\\
\widehat{Q}_{n}  &  =\frac{1}{n}\sum_{g=1}^{G}\boldsymbol{Z}_{g}^{\prime
}\boldsymbol{X}_{g},\\
\widehat{W}_{n}  &  =\frac{1}{n}\sum_{g=1}^{G}\boldsymbol{Z}_{g}^{\prime
}\boldsymbol{Z}_{g}.
\end{align*}
The variance estimator is%
\[
\widehat{V}_{n}=d_{n}\left(  \widehat{Q}_{n}^{\prime}\widehat{W}_{n}%
^{-1}\widehat{Q}_{n}\right)  ^{-1}\widehat{Q}_{n}^{\prime}\widehat{W}_{n}%
^{-1}\widehat{\Omega}_{n}\widehat{W}_{n}^{-1}\widehat{Q}_{n}\left(
\widehat{Q}_{n}^{\prime}\widehat{W}_{n}^{-1}\widehat{Q}_{n}\right)  ^{-1}.
\]
with $d_{n}$ a possible finite-sample degree-of-freedom adjustment. For
example, C. Hansen (2007) proposed $d_{n}=G/(G-1)$ for the regression case
(under homogeneous cluster sizes), and Stata sets%
\[
d_{n}=\left(  \frac{n-1}{n-k}\right)  \left(  \frac{G}{G-1}\right)
\]
for the OLS and 2SLS estimators under \textit{cluster} option.

\begin{theorem}
\label{2SLSd}Suppose that Assumption \ref{A2} holds for some $2\leq r\leq
s<\infty$, $Q_{n}$ has full rank $k$, $\lambda_{\min}(W_{n})\geq C>0$,
$\lambda_{\min}(\Omega_{n})\geq\lambda>0$, $\sup_{i}E\left\vert y_{i}%
\right\vert ^{2s}<\infty$, $\sup_{i}E\left\Vert \boldsymbol{x}_{i}\right\Vert
^{2s}<\infty$, and $\sup_{i}E\left\Vert \boldsymbol{z}_{i}\right\Vert
^{2s}<\infty$, and either

\begin{enumerate}
\item $(y_{i},\boldsymbol{x}_{i},\boldsymbol{z}_{i})$ have identical marginal
distributions; or

\item $r<s$;
\end{enumerate}

then, for any sequence of full-rank $k\times q$ matrices $R_{n}$, as
$n\rightarrow\infty$
\begin{equation}
\left(  R_{n}^{\prime}V_{n}R_{n}\right)  ^{-1/2}R_{n}^{\prime}\sqrt{n}\left(
\widehat{\boldsymbol{\beta}}-\boldsymbol{\beta}\right)  \xrightarrow {d}
N\left(  \mathbf{0},I_{q}\right)  , \label{2slsd1}%
\end{equation}
\begin{equation}
\left(  R_{n}^{\prime}V_{n}R_{n}\right)  ^{-1/2}R_{n}^{\prime}\widehat{V}
_{n}R_{n}\left(  R_{n}^{\prime}V_{n}R_{n}\right)  ^{-1/2}\xrightarrow{p}I_{q},
\label{2slsd2}%
\end{equation}
and
\begin{equation}
\left(  R_{n}^{\prime}\widehat{V}_{n}R_{n}\right)  ^{-1/2}R_{n}^{\prime}
\sqrt{n}\left(  \widehat{\boldsymbol{\beta}}-\boldsymbol{\beta}\right)
\xrightarrow {d}N\left(  \mathbf{0},I_{q}\right)  . \label{2slsd3}%
\end{equation}

\end{theorem}

The standard errors for $R_{n}^{\prime}\widehat{\boldsymbol{\beta}}$ can be
obtained by taking the square roots of the diagonal elements of $n^{-1}%
R_{n}^{\prime}\widehat{V}_{n}R_{n}$.

\section{(Pseudo) Maximum Likelihood}

Suppose that we observe a sequence of random vectors $X_{i}\in\mathbb{R}^{p}$,
$i=1,...,n$ with the same marginal distributions from a density
$f(x,\boldsymbol{\theta})$ where $\boldsymbol{\theta}\in\mathbb{R}^{k}$. Let
$\boldsymbol{X}_{g}=(X_{g1},...,X_{gn_{g}})^{\prime}$ be a $n_{g}\times p$
matrix for each cluster. For the observations in the cluster $g$, let
$f_{g}(\boldsymbol{X}_{g},\boldsymbol{\theta}_{0})$ be the joint density.
Since the observations within the same cluster need not be independent,
$f_{g}(\boldsymbol{X}_{g},\boldsymbol{\theta}_{0})\neq\prod_{i=1}^{n_{g}%
}f(X_{gi},\boldsymbol{\theta}_{0})$ in general. This also implies that
$f_{g}(\boldsymbol{X}_{g},\boldsymbol{\theta}_{0})\neq f_{h}(\boldsymbol{X}%
_{h},\boldsymbol{\theta}_{0})$ for $g\neq h$. Given specification of
$f_{g}(\boldsymbol{X}_{g},\boldsymbol{\theta}_{0})$, the maximum likelihood
estimator (MLE) can be obtained as the maximizer of
\[
\sum_{g=1}^{G}\log f_{g}(\boldsymbol{X}_{g},\boldsymbol{\theta}).
\]
However, the joint density $f_{g}(\boldsymbol{X}_{g},\boldsymbol{\theta})$ may
be difficult to specify in practice. A simpler alternative is to use a
pseudo-likelihood $\prod_{i=1}^{n_{g}}f(X_{gi},\boldsymbol{\theta}_{0})$ for
the joint density $f_{g}(\boldsymbol{X}_{g},\boldsymbol{\theta}_{0})$, and
specify the log likelihood function as
\[
L_{n}(\theta)=\sum_{g=1}^{G}\sum_{j=1}^{n_{g}}\log f(X_{gj},\boldsymbol{\theta
}).
\]
Define the pseudo-MLE as
\[
\widehat{\boldsymbol{\theta}}=\argmax_{\theta\in\Theta}L_{n}%
(\boldsymbol{\theta}).
\]
This estimator is also called the partial (or pooled) MLE (Wooldridge, 2010).

This estimator is the standard implementation of MLE under clustered
dependence. To our knowledge there is no existing distribution theory for this
standard estimator.

We first show consistency of $\widehat{\boldsymbol{\theta}}$. The following is
based on Theorem 2.1 of Newey and McFadden (1994).

\begin{theorem}
If Assumption \ref{A1} holds, \label{MLEc}

\begin{enumerate}
\item $X_{i}$ have identical marginal distributions with the density
$f(x,\boldsymbol{\theta}_{0})$ and $\boldsymbol{\theta}_{0}\in
\boldsymbol{\Theta}$, which is compact,

\item if $\boldsymbol{\theta}\neq\boldsymbol{\theta}_{0}$ then
$f(x,\boldsymbol{\theta})\neq f(x,\boldsymbol{\theta}_{0})$,

\item $E[\sup_{\boldsymbol{\theta} \in\Theta}|\log f(X_{i},\boldsymbol{\theta
})|]<\infty$,

\item for each $\boldsymbol{\theta} _{1},\boldsymbol{\theta} _{2}\in\Theta$,
\begin{equation}
\left\Vert \log f(x,\boldsymbol{\theta} _{1})-\log f(x,\boldsymbol{\theta}
_{2})\right\Vert \leq A(x)h\left(  \left\Vert \theta_{1}-\theta_{2}\right\Vert
\right) \nonumber
\end{equation}
where $h(u)\downarrow0$ as $u\downarrow0$ and $EA(X_{i})\leq C$,
\end{enumerate}

Then as $n\rightarrow\infty$, $\widehat{\boldsymbol{\theta}}%
\xrightarrow{p}\boldsymbol{\theta}_{0}.$
\end{theorem}

Next we show the asymptotic distribution. Define
\begin{align}
H_{n}(\boldsymbol{\theta})  &  = \frac{1}{n}\sum_{i=1}^{n}E\left[
\frac{\partial^{2}}{\partial\boldsymbol{\theta}\partial\boldsymbol{\theta
}^{\prime}}\log f(X_{i},\boldsymbol{\theta})\right]  ,\nonumber\\
\Omega_{n}(\boldsymbol{\theta})  &  = \frac{1}{n}\sum_{g=1}^{G}E\left(
\sum_{j=1}^{n_{g}} \frac{\partial}{\partial\boldsymbol{\theta}}\log
f(X_{gj},\boldsymbol{\theta})\right)  \left(  \sum_{j=1}^{n_{g}}
\frac{\partial}{\partial\boldsymbol{\theta}^{\prime}}\log f(X_{gj}%
,\boldsymbol{\theta})\right)  ,\nonumber\\
V_{n}  &  = H_{n}(\boldsymbol{\theta}_{0})^{-1}\Omega_{n}(\boldsymbol{\theta
}_{0})H_{n}(\boldsymbol{\theta}_{0})^{-1}.\nonumber
\end{align}

Define the sample versions
\begin{align}
\widehat{H}_{n}(\boldsymbol{\theta})  &  = \frac{1}{n}\sum_{i=1}^{n}%
\frac{\partial^{2}}{\partial\boldsymbol{\theta}\partial\boldsymbol{\theta
}^{\prime}}\log f(X_{i},\boldsymbol{\theta}),\nonumber\\
\widehat{\Omega}_{n}(\boldsymbol{\theta})  &  = \frac{1}{n}\sum_{g=1}%
^{G}\left(  \sum_{j=1}^{n_{g}} \frac{\partial}{\partial\boldsymbol{\theta}%
}\log f(X_{gj},\boldsymbol{\theta})\right)  \left(  \sum_{j=1}^{n_{g}}
\frac{\partial}{\partial\boldsymbol{\theta}^{\prime}}\log f(X_{gj}%
,\boldsymbol{\theta})\right)  .\nonumber
\end{align}
The variance estimator is
\[
\widehat{V}_{n} = \widehat{H}_{n}(\widehat{\boldsymbol{\theta}})^{-1}%
\widehat{\Omega}_{n}(\widehat{\boldsymbol{\theta}})\widehat{H}_{n}%
(\widehat{\boldsymbol{\theta}})^{-1}.
\]
Note that the information matrix equality does not hold because $\sum
_{j=1}^{n_{g}}\log f(X_{gj},\boldsymbol{\theta}_{0})\neq f_{g}(\boldsymbol{X}%
_{g},\boldsymbol{\theta}_{0})$ in general.

\begin{theorem}
\label{MLEd}In addition to the assumptions of Theorem \ref{MLEc}, Assumption
\ref{A2} holds with $r=2$,

\begin{enumerate}
\item $\boldsymbol{\theta}_{0}\in\text{interior}(\boldsymbol{\Theta})$,

\item for some neighborhood $\mathcal{N}$ of $\boldsymbol{\theta}_{0}$,

\begin{enumerate}
\item $f(x,\boldsymbol{\theta})$ is twice continuously differentiable and
$f(x,\boldsymbol{\theta})>0$,

\item $\int\sup_{\boldsymbol{\theta}\in\mathcal{N}}\left\Vert \frac{\partial
}{\partial\boldsymbol{\theta}}\log f(x,\boldsymbol{\theta})\right\Vert
dx<\infty$,

\item $E\left\|  \frac{\partial}{\partial\boldsymbol{\theta}}\log
f(X_{i},\boldsymbol{\theta})\right\|  ^{2}<\infty$,

\item $E\sup_{\boldsymbol{\theta} \in\mathcal{N} }\left\Vert \frac
{\partial^{2}}{\partial\boldsymbol{\theta}\partial\boldsymbol{\theta}^{\prime
}}\log f(X_{i},\boldsymbol{\theta})\right\Vert ^{2} <\infty$,

\item and for each $\boldsymbol{\theta}_{1},\boldsymbol{\theta}_{2}%
\in\mathcal{N}$,
\[
\left\Vert \frac{\partial^{2}}{\partial\boldsymbol{\theta}\partial
\boldsymbol{\theta}^{\prime}}\log f(x,\boldsymbol{\theta}_{1})-\frac
{\partial^{2}}{\partial\boldsymbol{\theta}\partial\boldsymbol{\theta}^{\prime
}}\log f(x,\boldsymbol{\theta}_{2})\right\Vert \leq A(x)h\left(  \left\Vert
\boldsymbol{\theta}_{1}-\boldsymbol{\theta}_{2}\right\Vert \right)
\]
where $h(u)\downarrow0$ as $u\downarrow0$ and $EA(X_{i})\leq C$,
\end{enumerate}

\item $\lambda_{\min}(H_{n}(\boldsymbol{\theta}_{0}))\geq C>0$,

\item $\lambda_{\min}(\Omega_{n}(\boldsymbol{\theta}_{0}))\geq\lambda>0$,
\end{enumerate}

then for any sequence of full-rank $k\times q$ matrices $R_{n}$, as
$n\rightarrow\infty$%
\begin{equation}
\left(  R_{n}^{\prime}V_{n}R_{n}\right)  ^{-1/2}R_{n}^{\prime}\sqrt{n}\left(
\widehat{\boldsymbol{\theta}}-\boldsymbol{\theta}_{0}\right)
\xrightarrow {d}N\left(  \mathbf{0},I_{q}\right)  , \label{mled1}%
\end{equation}%
\begin{equation}
\left(  R_{n}^{\prime}V_{n}R_{n}\right)  ^{-1/2}R_{n}^{\prime}\widehat{V}%
_{n}R_{n}\left(  R_{n}^{\prime}V_{n}R_{n}\right)  ^{-1/2}\xrightarrow{p}I_{q},
\label{mled2}%
\end{equation}
and%
\begin{equation}
\left(  R_{n}^{\prime}\widehat{V}_{n}R_{n}\right)  ^{-1/2}R_{n}^{\prime}%
\sqrt{n}\left(  \widehat{\boldsymbol{\theta}}-\boldsymbol{\theta}_{0}\right)
\xrightarrow {d}N\left(  \mathbf{0},I_{q}\right)  . \label{mled3}%
\end{equation}

\end{theorem}

The standard errors for $R_{n}^{\prime}\widehat{\boldsymbol{\beta}}$ can be
obtained by taking the square roots of the diagonal elements of $n^{-1}%
R_{n}^{\prime}\widehat{V}_{n}R_{n}$.

\section{Generalized Method of Moments}

Suppose that we observe a sequence of random vectors $X_{i}\in\mathbb{R}^{p}$,
$i=1,...,n$ from cluster sampling. A known moment function is given by
$m(X_{i},\boldsymbol{\theta})$ where $m(\cdot,\cdot)$ is $l\times1$ and
$\boldsymbol{\theta}$ is $k\times1$. Define the cluster sum as
\[
\widetilde{m}_{g}(\boldsymbol{\theta})=\sum_{j=1}^{n_{g}}m(X_{gj}%
,\boldsymbol{\theta}).
\]
An unconditional moment model in cluster notation is given by
\begin{equation}
E\widetilde{m}_{g}(\boldsymbol{\theta}_{0})=0. \label{mcc}%
\end{equation}
We assume that $\boldsymbol{\theta}_{0}$ is identified and $l>k$ so the moment
model is over-identified. Write the sample mean of the moment function as
\[
\overline{m}_{n}(\boldsymbol{\theta})=\frac{1}{n}\sum_{i=1}^{n}m(X_{i}%
,\boldsymbol{\theta}).
\]
Since \eqref{mcc} holds for all $g=1,...,G$, the usual unconditional moment
condition $E\overline{m}_{n}(\boldsymbol{\theta}_{0})=0$ follows. The
generalized method of moments (GMM) estimator is given by
\begin{equation}
\widehat{\boldsymbol{\theta}}=\argmin_{\boldsymbol{\theta}\in\Theta}%
n\cdot\overline{m}_{n}(\boldsymbol{\theta})^{\prime}\widehat{W}_{n}%
^{-1}\overline{m}_{n}(\boldsymbol{\theta}) \label{gmmcrit}%
\end{equation}
where $\widehat{W}_{n}^{-1}$ is an $l\times l$ positive definite weight
matrix, which may or may not depend on an estimated parameter. Typically, the
weight matrix is obtained by plugging in a preliminary consistent estimator,
$\widetilde{\boldsymbol{\theta}}$, so that $\widehat{W}_{n}^{-1}%
=\widehat{W}_{n}(\widetilde{\boldsymbol{\theta}})^{-1}$.

We consider two forms of GMM estimator. The first one is based on a
non-clustered weight matrix, which takes the form of
\begin{equation}
\widehat{W}_{n}(\boldsymbol{\theta})=\frac{1}{n}\sum_{i=1}^{n}v(X_{i}%
,\boldsymbol{\theta})v(X_{i},\boldsymbol{\theta})^{\prime} \label{convW}%
\end{equation}
for some $l\times1$ vector $v(x,\boldsymbol{\theta})$. This includes the
conventional one-step and two-step GMM estimators. For 2SLS, $v(X_{i}%
,\boldsymbol{\theta})=Z_{i}$ where $Z_{i}$ is an $l\times1$ vector of
instruments. The efficient two-step GMM uses $v(X_{i},\boldsymbol{\theta
})=m(X_{i},\boldsymbol{\theta})$ or $v(X_{i},\boldsymbol{\theta}%
)=m(X_{i},\boldsymbol{\theta})-\overline{m}_{n}(\boldsymbol{\theta})$. The
conventional efficient weight matrix, however, does not provide efficiency
anymore under cluster sampling because a weight matrix of the form of
\eqref{convW} is not consistent for the variance matrix of $\sqrt{n}%
(\overline{m}_{n}(\boldsymbol{\theta})-E\overline{m}_{n}(\boldsymbol{\theta
}))$ in general.

The second is based on the clustered efficient weight matrix, which leads to
the two-step efficient GMM under cluster sampling. The weight matrix takes the
form of
\begin{equation}
\widehat{W}_{n}(\boldsymbol{\theta})= \frac{1}{n}\sum_{g=1}^{G}\widetilde{m}%
_{g}(\boldsymbol{\theta})\widetilde{m}_{g}(\boldsymbol{\theta})^{\prime} -
\frac{1}{n}\sum_{g=1}^{G}n_{g}^{2}\overline{m}_{n}(\boldsymbol{\theta
})\overline{m}_{n}(\boldsymbol{\theta})^{\prime}. \label{cluWc}%
\end{equation}
Alternatively, the uncentered version of $\widehat{W}_{n}(\boldsymbol{\theta
})$ and $\widehat{\Omega}_{n}(\boldsymbol{\theta})$ can be used to obtain the
efficient two-step GMM estimator but the centered version is generally
recommended. For more discussion, see Hansen (2018).

Since we assume that the weight matrix depends on a consistent preliminary
estimator, we exclude the continuously updating (CU) GMM estimator in our
analysis. Whenever possible, we omit the dependence of the weight matrices on
$\widetilde{\boldsymbol{\theta}}$ and write $\widehat{W}_{n} = \widehat{W}%
_{n}(\widetilde{\boldsymbol{\theta}})$. Define $W_{n} = E\widehat{W}%
_{n}(\boldsymbol{\theta}_{0})$.

We first show consistency of the GMM estimator. The following is based on
Theorem 2.1 of Newey and McFadden (1994).

\begin{theorem}
\label{GMMc} If Assumption \ref{A1} holds,

\begin{enumerate}
\item $\Theta$ is compact,

\item $\boldsymbol{\theta}_{0}$ is the unique solution to $E\overline{m}%
_{n}(\boldsymbol{\theta}) = 0$,

\item for each $\boldsymbol{\theta}\in\boldsymbol{\Theta}$, either $X_{i}$
have identical marginal distributions with $E\left\Vert m(X_{i}%
,\boldsymbol{\theta})\right\Vert <\infty$, or $\sup_{i}E\left\Vert
m(X_{i},\boldsymbol{\theta})\right\Vert ^{r}<\infty$ for some $r>1$,

\item for each $\boldsymbol{\theta} _{1},\boldsymbol{\theta} _{2}\in\Theta$
\begin{equation}
\left\Vert m(x,\boldsymbol{\theta} _{1})-m(x,\boldsymbol{\theta}
_{2})\right\Vert \leq A(x)h\left(  \left\Vert \boldsymbol{\theta}
_{1}-\boldsymbol{\theta} _{2}\right\Vert \right) \nonumber
\end{equation}
where $h(u)\downarrow0$ as $u\downarrow0$ and $EA(X_{i})\leq C$,

\item $\lambda_{\min}(W_{n})\geq C>0$,

\item $\widehat{W}_{n}^{-1}-W_{n}^{-1}\xrightarrow{p}0$,
\end{enumerate}

then as $n\rightarrow\infty$, $\widehat{\boldsymbol{\theta}}%
\xrightarrow{p}\boldsymbol{\theta}_{0}.$
\end{theorem}

Primitive conditions under which Condition 6 of Theorem \ref{GMMc} holds can
be found given the choice of the weight matrix. For simplicity, we assume that
if the conventional weight matrix is used then either $v(X_{i},\boldsymbol{
\theta}) =m(X_{i},\boldsymbol{\theta})$ or $v(X_{i},\boldsymbol{\theta}) =
m(X_{i},\boldsymbol{\theta})-\overline{m}_{n}(\boldsymbol{\theta})$. If the
clustered weight matrix is used then it takes the form of \eqref{cluWc}. The
conditions of Theorem \ref{GMMd} are sufficient for Condition 6 of Theorem
\ref{GMMc} to hold.

To show the asymptotic distribution of the GMM estimator, define
\begin{align}
Q_{n}(\boldsymbol{\theta})  &  =\frac{1}{n}\sum_{i=1}^{n}E\left[
\frac{\partial}{\partial\boldsymbol{\theta}^{\prime}}m(X_{i}%
,\boldsymbol{\theta})\right]  ,\nonumber\\
\Omega_{n}(\boldsymbol{\theta})  &  =\frac{1}{n}\sum_{g=1}^{G}E\widetilde{m}%
_{g}(\boldsymbol{\theta})\widetilde{m}_{g}(\boldsymbol{\theta})^{\prime
},\nonumber\\
V_{n}  &  =(Q_{n}^{\prime}W_{n}^{-1}Q_{n})^{-1}Q_{n}^{\prime}W_{n}^{-1}%
\Omega_{n}W_{n}^{-1}Q_{n}(Q_{n}^{\prime}W_{n}^{-1}Q_{n})^{-1},\nonumber
\end{align}
where $Q_{n}=Q_{n}(\boldsymbol{\theta}_{0})$ and $\Omega_{n}=\Omega
_{n}(\boldsymbol{\theta}_{0})$. If the clustered efficient weight matrix
\eqref{cluWc} is used, then the asymptotic variance matrix simplifies to
\[
V_{n}=(Q_{n}^{\prime}\Omega_{n}^{-1}Q_{n})^{-1}.
\]
Define the sample versions as
\begin{align}
\widehat{Q}_{n}(\boldsymbol{\theta})  &  =\frac{1}{n}\sum_{i=1}^{n}%
\frac{\partial}{\partial\boldsymbol{\theta}^{\prime}}m(X_{i}%
,\boldsymbol{\theta}),\nonumber\\
\widehat{\Omega}_{n}(\boldsymbol{\theta})  &  =\frac{1}{n}\sum_{g=1}%
^{G}\widetilde{m}_{g}(\boldsymbol{\theta})\widetilde{m}_{g}(\boldsymbol{\theta
})^{\prime}-\frac{1}{n}\sum_{g=1}^{G}n_{g}^{2}\overline{m}_{n}%
(\boldsymbol{\theta})\overline{m}_{n}(\boldsymbol{\theta})^{\prime}\nonumber
\end{align}
and let $\widehat{Q}_{n}=\widehat{Q}_{n}(\widehat{\boldsymbol{\theta}})$ and
$\widehat{\Omega}_{n}=\widehat{\Omega}_{n}(\widehat{\boldsymbol{\theta}})$.
The variance estimator is
\[
\widehat{V}_{n}=(\widehat{Q}_{n}^{\prime}\widehat{W}_{n}^{-1}\widehat{Q}%
_{n})^{-1}\widehat{Q}_{n}^{\prime}\widehat{W}_{n}^{-1}\widehat{\Omega}%
_{n}\widehat{W}_{n}^{-1}\widehat{Q}_{n}(\widehat{Q}_{n}^{\prime}%
\widehat{W}_{n}^{-1}\widehat{Q}_{n})^{-1},
\]
if $\widehat{W}_{n}$ is given by \eqref{convW} and
\[
\widehat{V}_{n}=(\widehat{Q}_{n}^{\prime}\widehat{\Omega}_{n}^{-1}%
\widehat{Q}_{n})^{-1},
\]
if $\widehat{W}_{n}$ is given by \eqref{cluWc}, i.e., $\widehat{W}%
_{n}=\widehat{\Omega}_{n}$.

The over-identifying restrictions test (the J test, hereinafter) is a test
based on the GMM criterion to test whether the moment model is correctly
specified or not, i.e., $E\widetilde{m}_{g}(\boldsymbol{\theta}_{0})=0$. An
implication of cluster sampling is that the conventional J test statistic will
not have a standard chi-square asymptotic distribution because the
conventional efficient weight matrix is not consistent for the inverse of the
variance matrix of the moment function. The GMM criterion \eqref{gmmcrit}
based on the clustered efficient weight matrix \eqref{cluWc} evaluated at the
estimator is the robust J test statistic. Define
\[
J_{n}(\widehat{\boldsymbol{\theta}})=n\cdot\overline{m}_{n}%
(\widehat{\boldsymbol{\theta}})^{\prime}\widehat{W}_{n}^{-1}\overline{m}%
_{n}(\widehat{\boldsymbol{\theta}}).
\]

\begin{theorem}
\label{GMMd} In addition to the assumptions of Theorem \ref{GMMc}, if
Assumption \ref{A2} holds with $r=2$,

\begin{enumerate}
\item $\boldsymbol{\theta}_{0}\in\text{interior}(\boldsymbol{\Theta})$,

\item for some neighborhood $\mathcal{N}$ of $\boldsymbol{\theta}_{0}$,

\begin{enumerate}
\item $m(X_{i},\boldsymbol{\theta})$ is continuously differentiable with
probability approaching one,

\item either $X_{i}$ have identical marginal distributions with $E\sup
_{\boldsymbol{\theta} \in\mathcal{N}}\left\Vert m(X_{i},\boldsymbol{\theta
})\right\Vert ^{2}<\infty$;\newline or $E\sup_{i}\sup_{\boldsymbol{\theta}
\in\mathcal{N}}\left\Vert m(X_{i},\boldsymbol{\theta})\right\Vert ^{r}<\infty$
for some $r>2$,

\item $E\sup_{i}\sup_{\boldsymbol{\theta} \in\mathcal{N}}\left\Vert
\frac{\partial}{\partial\boldsymbol{\theta}^{\prime}}m(X_{i}%
,\boldsymbol{\theta})\right\Vert ^{2}<\infty$

\item for each $\boldsymbol{\theta} _{1},\boldsymbol{\theta} _{2}%
\in\mathcal{N} $
\begin{equation}
\left\Vert \frac{\partial}{\partial\boldsymbol{\theta}}m(x,\boldsymbol{\theta}
_{1})-\frac{\partial}{\partial\boldsymbol{\theta}}m(x,\boldsymbol{\theta}
_{2})\right\Vert \leq A(x)h\left(  \left\Vert \boldsymbol{\theta} _{1}-
\boldsymbol{\theta} _{2}\right\Vert \right) \nonumber
\end{equation}
where $h(u)\downarrow0$ as $u\downarrow0$ and $\sup_{i}EA(X_{i})\leq C$,
\end{enumerate}

\item $\lambda_{\min}(W_{n}(\boldsymbol{\theta}_{0}))\geq C >0$,

\item $\lambda_{\min}(\Omega_{n}(\boldsymbol{\theta}_{0}))\geq\lambda>0$,

\item $Q_{n}$ is full column rank,
\end{enumerate}

then for any sequence of full-rank $k\times q$ matrices $R_{n}$, as
$n\rightarrow\infty$%
\begin{equation}
\left(  R_{n}^{\prime}V_{n}R_{n}\right)  ^{-1/2}R_{n}^{\prime}\sqrt{n}\left(
\widehat{\boldsymbol{\theta}}-\boldsymbol{\theta}_{0}\right)
\xrightarrow {d}N\left(  \mathbf{0},I_{q}\right)  , \label{gmmd1}%
\end{equation}%
\begin{equation}
\left(  R_{n}^{\prime}V_{n}R_{n}\right)  ^{-1/2}R_{n}^{\prime}\widehat{V}%
_{n}R_{n}\left(  R_{n}^{\prime}V_{n}R_{n}\right)  ^{-1/2}\xrightarrow{p}I_{q},
\label{gmmd2}%
\end{equation}
\begin{equation}
\left(  R_{n}^{\prime}\widehat{V}_{n}R_{n}\right)  ^{-1/2}R_{n}^{\prime}%
\sqrt{n}\left(  \widehat{\boldsymbol{\theta}}-\boldsymbol{\theta}_{0}\right)
\xrightarrow {d}N\left(  \mathbf{0},I_{q}\right)  , \label{gmmd3}%
\end{equation}
and
\begin{equation}
J_{n}(\widehat{\boldsymbol{\theta} })\xrightarrow{d}\chi^{2}_{l-k}.
\label{gmmd4}%
\end{equation}

\end{theorem}

The standard errors for $R_{n}^{\prime}\widehat{\boldsymbol{\beta}}$ can be
obtained by taking the square roots of the diagonal elements of $n^{-1}%
R_{n}^{\prime}\widehat{V}_{n}R_{n}$.

\newpage

\section{Appendix}

We start with a useful technical result which states that if random variables
are uniformly integrable then so are their cluster averages, regardless of
their joint dependence.

\begin{lemma}
\label{uitheorem}For random vectors $X_{i}$ set $\widetilde{X}_{m}=\sum
_{i=1}^{m}X_{i}$. For $r\geq1$, if
\begin{equation}
\lim_{B\rightarrow\infty}\sup_{i}E\left(  \left\Vert X_{i}\right\Vert
^{r}1\left(  \left\Vert X_{i}\right\Vert >B\right)  \right)  =0, \label{supx}%
\end{equation}
then
\begin{align}
&  \lim_{B\rightarrow\infty}\sup_{m}E\left(  \left\Vert m^{-1}\widetilde{X}%
_{m}\right\Vert ^{r}1\left(  \left\Vert m^{-1}\widetilde{X}_{m}\right\Vert
>B\right)  \right)  =0 . \label{supxb}%
\end{align}

\end{lemma}

\noindent\textbf{Proof of Lemma \ref{uitheorem}:} The proof is based on the
proof of Theorem 1 of Etemadi (2006). Equation (\ref{supx}) implies that
$\sup_{i}E\left\Vert X_{i}\right\Vert ^{r}\leq C$ for some $C<\infty$. By the
$C_{r}$ inequality%
\begin{equation}
\left\Vert m^{-1}\widetilde{X}_{m}\right\Vert ^{r}=\frac{1}{m^{r}}\left\Vert
\sum_{i=1}^{m}X_{i}\right\Vert ^{r}\leq\frac{1}{m}\sum_{i=1}^{m}\left\Vert
X_{i}\right\Vert ^{r} \label{xr}%
\end{equation}
and hence
\begin{equation}
E\left\Vert m^{-1}\widetilde{X}_{m}\right\Vert ^{r}\leq C. \label{exr}%
\end{equation}

Fix $\varepsilon>0$. Find $B\geq(C/\varepsilon)^{2/r}$ sufficiently large such
that%
\begin{equation}
\sup_{i}E\left(  \left\Vert X_{i}\right\Vert ^{r}1\left(  \left\Vert
X_{i}\right\Vert >\sqrt{B}\right)  \right)  \leq\varepsilon, \label{supx2}%
\end{equation}
which is feasible under (\ref{supx}). Using (\ref{xr}),%
\begin{align*}
&  E\left(  \left\Vert m^{-1}\widetilde{X}_{m}\right\Vert ^{r}1\left(
\left\Vert m^{-1}\widetilde{X}_{m}\right\Vert >B\right)  \right) \\
&  \leq\frac{1}{m}\sum_{i=1}^{m}E\left(  \left\Vert X_{i}\right\Vert
^{r}1\left(  \left\Vert m^{-1}\widetilde{X}_{m}\right\Vert >B\right)  \right)
\\
&  =\frac{1}{m}\sum_{i=1}^{m}E\left(  \left\Vert X_{i}\right\Vert ^{r}1\left(
\left\Vert m^{-1}\widetilde{X}_{m}\right\Vert >B\right)  1\left(  \left\Vert
X_{i}\right\Vert >\sqrt{B}\right)  \right) \\
&  +\frac{1}{m}\sum_{i=1}^{m}E\left(  \left\Vert X_{i}\right\Vert ^{r}1\left(
\left\Vert m^{-1}\widetilde{X}_{m}\right\Vert >B\right)  1\left(  \left\Vert
X_{i}\right\Vert \leq\sqrt{B}\right)  \right) \\
&  \leq\frac{1}{m}\sum_{i=1}^{m}E\left(  \left\Vert X_{i}\right\Vert
^{r}1\left(  \left\Vert X_{i}\right\Vert >\sqrt{B}\right)  \right)
+B^{r/2}E1\left(  \left\Vert m^{-1}\widetilde{X}_{m}\right\Vert >B\right) \\
&  \leq\varepsilon+\frac{E\left\Vert m^{-1}\widetilde{X}_{m}\right\Vert ^{r}%
}{B^{r/2}}\\
&  \leq2\varepsilon
\end{align*}
by (\ref{supx2}), Markov's inequality, (\ref{exr}), and $B^{r/2}\geq
C/\varepsilon$. Since $\varepsilon$ is arbitrary this implies (\ref{supxb}).
\qquad$\blacksquare$

\bigskip

The next Lemma is useful for establishing the WLLN and CLT for the vectorized
clustered second moments.

\begin{lemma}
\label{uitheorem2}For random vectors $X_{i}$ set $\widetilde{X} _{m}%
=\sum_{i=1}^{m}X_{i}$ and $\widetilde{f}_{m} =\widetilde{X}_{m}\otimes
\widetilde{X}_{m}$ or $\widetilde{f}_{m} =\left(  \widetilde{X}_{m}%
-m\overline{X}_{n}\right)  \otimes\left(  \widetilde{X}_{m}-m\overline{X}%
_{n}\right)  $ where $\overline{X}_{n} = n^{-1}\sum_{i=1}^{n}X_{i}$. For
$r\geq2$, if (\ref{supx}) holds then
\begin{align}
&  \lim_{B\rightarrow\infty}\sup_{m}E\left(  \left\Vert m^{-2}\left(
\widetilde{f}_{m}-E\widetilde{f}_{m}\right)  \right\Vert ^{r/2}1\left(
\left\Vert m^{-2}\left(  \widetilde{f}_{m}-E\widetilde{f}_{m}\right)
\right\Vert >B\right)  \right)  =0 . \label{supxxb}%
\end{align}

\end{lemma}

\noindent\textbf{Proof of Lemma \ref{uitheorem2}:} The proof proceeds similar
to that of Lemma \ref{uitheorem}. First consider $\widetilde{f}_{m}%
=\widetilde{X}_{m}\otimes\widetilde{X}_{m}$. By the triangle inequality, the
$C_{r}$ inequality, the fact that $\Vert\widetilde{X}_{m}\otimes
\widetilde{X}_{m}\Vert^{r/2}=\Vert\widetilde{X}_{m}\Vert^{r}$, and
(\ref{exr}),
\begin{align}
\left\Vert m^{-2}\left(  \widetilde{f}_{m}-E\widetilde{f}_{m}\right)
\right\Vert ^{r/2}  &  \leq\left(  \left\Vert m^{-2}\widetilde{f}%
_{m}\right\Vert +\left\Vert m^{-2}E\widetilde{f}_{m}\right\Vert \right)
^{r/2}\nonumber\\
&  \leq2^{r/2-1}\left(  \left\Vert m^{-2}\widetilde{f}_{m}\right\Vert
^{r/2}+E\left\Vert m^{-2}\widetilde{f}_{m}\right\Vert ^{r/2}\right)
\nonumber\\
&  \leq2^{r/2-1}\left(  \left\Vert m^{-1}\widetilde{X}_{m}\right\Vert
^{r}+E\left\Vert m^{-1}\widetilde{X}_{m}\right\Vert ^{r}\right) \nonumber\\
&  \leq2^{r/2-1}\left(  \left\Vert m^{-1}\widetilde{X}_{m}\right\Vert
^{r}+C\right)  . \label{mf2}%
\end{align}

Fix $\varepsilon>0$. Find $B\geq\left(  2^{r-2}C(1+\sqrt{1+2^{3-r}\varepsilon
})/\varepsilon\right)  ^{4/r}$ sufficiently large such that%
\begin{equation}
\sup_{i}E\left(  \left\Vert X_{i}\right\Vert ^{r}1\left(  \left\Vert
X_{i}\right\Vert >B^{1/4}\right)  \right)  \leq\frac{\varepsilon}{2^{r/2-1}} ,
\label{supx4}%
\end{equation}
which is feasible under (\ref{supx}). Using (\ref{mf2}) and (\ref{xr}),%
\begin{align*}
&  E\left(  \left\Vert m^{-2}\left(  \widetilde{f}_{m}-E\widetilde{f}%
_{m}\right)  \right\Vert ^{r/2}1\left(  \left\Vert m^{-2}\left(
\widetilde{f}_{m}-E\widetilde{f}_{m}\right)  \right\Vert >B\right)  \right) \\
&  \leq2^{r/2-1}E\left(  \left(  \left\Vert m^{-1}\widetilde{X}_{m}\right\Vert
^{r} + C\right)  1\left(  \left\Vert m^{-2}\left(  \widetilde{f}%
_{m}-E\widetilde{f}_{m}\right)  \right\Vert >B\right)  \right) \\
&  =2^{r/2-1}\frac{1}{m}\sum_{i=1}^{m}E\left(  \left\Vert X_{i}\right\Vert
^{r}1\left(  \left\Vert m^{-2}\left(  \widetilde{f}_{m}-E\widetilde{f}%
_{m}\right)  \right\Vert >B\right)  1\left(  \left\Vert X_{i}\right\Vert
>B^{1/4}\right)  \right) \\
&  +2^{r/2-1}\frac{1}{m}\sum_{i=1}^{m}E\left(  \left\Vert X_{i}\right\Vert
^{r}1\left(  \left\Vert m^{-2}\left(  \widetilde{f}_{m}-E\widetilde{f}%
_{m}\right)  \right\Vert >B\right)  1\left(  \left\Vert X_{i}\right\Vert \leq
B^{1/4}\right)  \right) \\
&  + 2^{r/2-1}C E\left(  1\left(  \left\Vert m^{-2}\left(  \widetilde{f}%
_{m}-E\widetilde{f}_{m}\right)  \right\Vert >B\right)  \right) \\
&  \leq2^{r/2-1}\frac{1}{m}\sum_{i=1}^{m}E\left(  \left\Vert X_{i}\right\Vert
^{r}1\left(  \left\Vert X_{i}\right\Vert >B^{1/4}\right)  \right) \\
&  +2^{r/2-1}\left(  B^{r/4} + C\right)  E\left(  1\left(  \left\Vert
m^{-2}\left(  \widetilde{f}_{m}-E\widetilde{f}_{m}\right)  \right\Vert
>B\right)  \right) \\
&  \leq\varepsilon+2^{r/2-1}\left(  B^{r/4} + C\right)  \frac{E\left\Vert
m^{-2}\left(  \widetilde{f}_{m}-E\widetilde{f}_{m}\right)  \right\Vert ^{r/2}
}{B^{r/2}}\\
&  \leq2\varepsilon
\end{align*}
by (\ref{supx4}), Markov's inequality, (\ref{exr}), and $2^{r-1}%
(B^{r/4}+C)C/B^{r/2} \leq\varepsilon$ using the discriminant. Since
$\varepsilon$ is arbitrary this implies (\ref{supxxb}).

Now consider $\widetilde{f}_{m}=\left(  \widetilde{X}_{m}-m\overline{X}%
_{n}\right)  \otimes\left(  \widetilde{X}_{m}-m\overline{X}_{n}\right)  $. By
Minkowski's inequality, the $C_{r}$ inequality, (\ref{xr}), and (\ref{exr}),
\begin{align*}
E\left\Vert m^{-1}\left(  \widetilde{X}_{m}-m\overline{X}_{n}\right)
\right\Vert ^{r}  &  =E\left\Vert m^{-1}\sum_{i=1}^{m}X_{i}-n^{-1}\sum
_{i=1}^{n}X_{i}\right\Vert ^{r}\\
&  \leq E\left(  \left\Vert m^{-1}\sum_{i=1}^{m}X_{i}\right\Vert +\left\Vert
n^{-1}\sum_{i=1}^{n}X_{i}\right\Vert \right)  ^{r}\\
&  \leq2^{r}C
\end{align*}
and
\begin{align*}
\left\Vert m^{-2}\left(  \widetilde{f}_{m}-E\widetilde{f}_{m}\right)
\right\Vert ^{r/2}  &  \leq\left(  \left\Vert m^{-2}\widetilde{f}%
_{m}\right\Vert +\left\Vert m^{-2}E\widetilde{f}_{m}\right\Vert \right)
^{r/2}\\
&  \leq2^{r/2-1}\left(  \left\Vert m^{-1}\left(  \widetilde{X}_{m}%
-m\overline{X}_{n}\right)  \right\Vert ^{r}+E\left\Vert m^{-1}\left(
\widetilde{X}_{m}-m\overline{X}_{n}\right)  \right\Vert ^{r}\right) \\
&  \leq2^{3r/2-1}\left(  2^{-1}\left(  m^{-1}\sum_{i=1}^{m}\left\Vert
X\right\Vert ^{r}+n^{-1}\sum_{i=1}^{n}\left\Vert X_{i}\right\Vert ^{r}\right)
+C\right)  .
\end{align*}
Given $\varepsilon$, find $B\geq\left(  2^{3r-2}C(1+\sqrt{1+2^{3(1-r)}%
\varepsilon})/\varepsilon\right)  ^{4/r}$ sufficiently large such that%
\[
\sup_{i}E\left(  \left\Vert X_{i}\right\Vert ^{r}1\left(  \left\Vert
X_{i}\right\Vert >B^{1/4}\right)  \right)  \leq\frac{\varepsilon}{2^{3r/2-1}%
},
\]
and proceed as above to show (\ref{supxxb}). This completes the proof.
\qquad$\blacksquare$

\bigskip

\noindent\textbf{Proof of Theorem \ref{wlln}:} Without loss of generality
assume $EX_{i}=0$. Fix $\varepsilon>0$. Pick $B$ sufficiently large so that%
\begin{equation}
\sup_{g}E\left\Vert \left(  n_{g}^{-1}\widetilde{X}_{g}1\left(  \left\Vert
n_{g}^{-1}\widetilde{X}_{g}\right\Vert >B\right)  \right)  -E\left(
n_{g}^{-1}\widetilde{X}_{g}1\left(  \left\Vert n_{g}^{-1}\widetilde{X}%
_{g}\right\Vert >B\right)  \right)  \right\Vert \leq\varepsilon\label{ui}%
\end{equation}
which is feasible by Lemma \ref{uitheorem} with $r=1$ under (\ref{ui1}). Using
the triangle inequality, Jensen's inequality and (\ref{ui}),
\begin{align*}
&  E\left\Vert \overline{X}_{n}\right\Vert =E\left\Vert \frac{1}{n}\sum
_{g=1}^{G}\widetilde{X}_{g}\right\Vert \\
&  \leq E\left\Vert \frac{1}{n}\sum_{g=1}^{G}\left(  \widetilde{X}_{g}1\left(
\left\Vert n_{g}^{-1}\widetilde{X}_{g}\right\Vert \leq B\right)  -E\left(
\widetilde{X}_{g}1\left(  \left\Vert n_{g}^{-1}\widetilde{X}_{g}\right\Vert
\leq B\right)  \right)  \right)  \right\Vert \\
&  +\frac{1}{n}\sum_{g=1}^{G}E\left\Vert \left(  \widetilde{X}_{g}1\left(
\left\Vert n_{g}^{-1}\widetilde{X}_{g}\right\Vert >B\right)  -E\left(
\widetilde{X}_{g}1\left(  \left\Vert n_{g}^{-1}\widetilde{X}_{g}\right\Vert
>B\right)  \right)  \right)  \right\Vert \\
&  \leq\left(  E\left\Vert \frac{1}{n}\sum_{g=1}^{G}\left(  \widetilde{X}%
_{g}1\left(  \left\Vert n_{g}^{-1}\widetilde{X}_{g}\right\Vert \leq B\right)
-E\left(  \widetilde{X}_{g}1\left(  \left\Vert n_{g}^{-1}\widetilde{X}%
_{g}\right\Vert \leq B\right)  \right)  \right)  \right\Vert ^{2}\right)
^{1/2}+\frac{1}{n}\sum_{g=1}^{G}n_{g}\varepsilon\\
&  =\left(  \frac{1}{n^{2}}\sum_{g=1}^{G}E\left\Vert \widetilde{X}_{g}1\left(
\left\Vert n_{g}^{-1}\widetilde{X}_{g}\right\Vert \leq B\right)  -E\left(
\widetilde{X}_{g}1\left(  \left\Vert n_{g}^{-1}\widetilde{X}_{g}\right\Vert
\leq B\right)  \right)  \right\Vert ^{2}\right)  ^{1/2}+\varepsilon\\
&  \leq\left(  \frac{4B^{2}}{n^{2}}\sum_{g=1}^{G}n_{g}^{2}\right)
^{1/2}+\varepsilon\\
&  \leq o(1)+\varepsilon.
\end{align*}
The equality uses the assumption that the clusters are independent and thus
uncorrelated and the fact $\sum_{g=1}^{G}n_{g}=n$. The third inequality uses
the bound
\[
\left\Vert \widetilde{X}_{g}1\left(  \left\Vert n_{g}^{-1}\widetilde{X}%
_{g}\right\Vert \leq B\right)  -E\left(  \widetilde{X}_{g}1\left(  \left\Vert
n_{g}^{-1}\widetilde{X}_{g}\right\Vert \leq B\right)  \right)  \right\Vert
\leq2Bn_{g}.
\]
The fourth inequality is (\ref{n2n}). Since $\varepsilon$ is arbitrary,
$E\left\Vert \overline{X}_{n}\right\Vert \rightarrow0$. By Markov's
inequality,\ (\ref{wlln1}) follows.\qquad$\blacksquare$

\bigskip

\noindent\textbf{Proof of Theorem \ref{clt}:} Without loss of generality we
assume $EX_{i}=0.$ Note that%
\[
\Omega_{n}^{-1/2}\sqrt{n}\overline{X}_{n}=\Omega_{n}^{-1/2}\sum_{g=1}%
^{G}n^{-1/2}\widetilde{X}_{g}%
\]

We apply the multivariate Lindeberg-Feller central limit theorem (e.g. Hansen
(2018) Theorem 6.15) since $\widetilde{X}_{g}$ are independent but not
identically distributed. A sufficient condition for the CLT (\ref{clt1}) is
that for all $\varepsilon>0$%
\begin{equation}
\frac{1}{n\lambda_{n}}\sum_{g=1}^{G}E\left(  \left\Vert \widetilde{X}%
_{g}\right\Vert ^{2}1\left(  \left\Vert \widetilde{X}_{g}\right\Vert ^{2}\geq
n\lambda_{n}\varepsilon\right)  \right)  \rightarrow0 \label{lind1}%
\end{equation}
as $n\rightarrow\infty$. \ 

Fix $\varepsilon>0$ and $\delta>0$. Pick $B$ sufficiently large so that%
\begin{equation}
\sup_{g}E\left(  \left\Vert n_{g}^{-1}\widetilde{X}_{g}\right\Vert
^{r}1\left(  \left\Vert n_{g}^{-1}\widetilde{X}_{g}\right\Vert >B\right)
\right)  \leq\frac{\delta\varepsilon^{r/2-1}}{C^{r/2}}. \label{cltb1}%
\end{equation}
which is feasible by Lemma \ref{uitheorem} under (\ref{uir}). Pick $n$ large
enough so that%
\begin{equation}
\max_{g\leq G}\frac{n_{g}}{\left(  n\lambda_{n}\varepsilon\right)  ^{1/2}}%
\leq\frac{1}{B} \label{cltb2}%
\end{equation}
which is feasible by (\ref{n2}). Thus%
\begin{align}
&  \frac{1}{n\lambda_{n}}\sum_{g=1}^{G}E\left(  \left\Vert \widetilde{X}%
_{g}\right\Vert ^{2}1\left(  \left\Vert \widetilde{X}_{g}\right\Vert ^{2}\geq
n\lambda_{n}\varepsilon\right)  \right) \label{lind3}\\
&  =\frac{1}{n\lambda_{n}}\sum_{g=1}^{G}E\left(  \frac{\left\Vert
\widetilde{X}_{g}\right\Vert ^{r}}{\left\Vert \widetilde{X}_{g}\right\Vert
^{r-2}}1\left(  \left\Vert \widetilde{X}_{g}\right\Vert \geq\left(
n\lambda_{n}\varepsilon\right)  ^{1/2}\right)  \right) \nonumber\\
&  \leq\frac{1}{n\lambda_{n}\left(  n\lambda_{n}\varepsilon\right)
^{(r-2)/2}}\sum_{g=1}^{G}E\left(  \left\Vert \widetilde{X}_{g}\right\Vert
^{r}1\left(  \left\Vert \widetilde{X}_{g}\right\Vert \geq\left(  n\lambda
_{n}\varepsilon\right)  ^{1/2}\right)  \right) \nonumber\\
&  \leq\frac{1}{\varepsilon^{r/2-1}\left(  n\lambda_{n}\right)  ^{r/2}}%
\sum_{g=1}^{G}n_{g}^{r}E\left(  \left\Vert n_{g}^{-1}\widetilde{X}%
_{g}\right\Vert ^{r}1\left(  \left\Vert n_{g}^{-1}\widetilde{X}_{g}\right\Vert
\geq B\right)  \right) \nonumber\\
&  \leq\frac{\delta}{C^{r/2}}\frac{\sum_{g=1}^{G}n_{g}^{r}}{\left(
n\lambda_{n}\right)  ^{r/2}}\nonumber\\
&  \leq\delta.\nonumber
\end{align}
The second inequality is (\ref{cltb2}), the third is (\ref{cltb1}), and the
final is\ (\ref{n1}). Since $\varepsilon$ and $\delta$ are arbitrary we have
established (\ref{lind1}) and hence (\ref{clt1}).\qquad$\blacksquare$

\bigskip

\noindent\textbf{Proof of Theorem \ref{cov}:} Fix $\delta>0$. Set
$\varepsilon=\delta^{2}/4p$. Define $\widetilde{X}_{g}^{\ast}=\Omega
_{n}^{-1/2}\widetilde{X}_{g}$ and $\widetilde{Y}_{g}=\widetilde{X}_{g}^{\ast
}1\left(  \left\Vert \widetilde{X}_{g}^{\ast}\right\Vert ^{2}\leq
n\varepsilon\right)  $. Then%
\begin{align*}
\widetilde{\Omega}_{n}^{\ast}  &  =\frac{1}{n}\sum_{g=1}^{G}\widetilde{X}%
_{g}^{\ast}\widetilde{X}_{g}^{\ast\prime}\\
&  =\frac{1}{n}\sum_{g=1}^{G}\widetilde{Y}_{g}\widetilde{Y}_{g}^{\prime}%
+\frac{1}{n}\sum_{g=1}^{G}\widetilde{X}_{g}^{\ast}\widetilde{X}_{g}%
^{\ast\prime}1\left(  \left\Vert \widetilde{X}_{g}^{\ast}\right\Vert
^{2}>n\varepsilon\right)  .
\end{align*}
By the triangle inequality,
\begin{align}
E\left\Vert \widetilde{\Omega}_{n}^{\ast}-I_{p}\right\Vert  &  \leq\frac{1}%
{n}E\left\Vert \sum_{g=1}^{G}\left(  \widetilde{Y}_{g}\widetilde{Y}%
_{g}^{\prime}-E\left(  \widetilde{Y}_{g}\widetilde{Y}_{g}^{\prime}\right)
\right)  \right\Vert \label{evv1}\\
&  +\frac{2}{n}\sum_{g=1}^{G}E\left(  \left\Vert \widetilde{X}_{g}^{\ast
}\right\Vert ^{2}1\left(  \left\Vert \widetilde{X}_{g}^{\ast}\right\Vert
^{2}>n\varepsilon\right)  \right)  . \label{evv2}%
\end{align}

An argument similar to (\ref{lind3}) shows that for $n$ sufficiently large
(\ref{evv2}) is bounded by $2\delta$. We now consider (\ref{evv1}).

Using Jensen's inequality, the assumption that the clusters are independent
and thus uncorrelated, and the triangle inequality, (\ref{evv1}) is bounded
by
\begin{align}
\frac{1}{n}\left(  E\left\Vert \sum_{g=1}^{G}\left(  \widetilde{Y}%
_{g}\widetilde{Y}_{g}^{\prime}-E\left(  \widetilde{Y}_{g}\widetilde{Y}%
_{g}^{\prime}\right)  \right)  \right\Vert ^{2}\right)  ^{1/2} &  =\frac{1}%
{n}\left(  \sum_{g=1}^{G}E\left\Vert \widetilde{Y}_{g}\widetilde{Y}%
_{g}^{\prime}-E\left(  \widetilde{Y}_{g}\widetilde{Y}_{g}^{\prime}\right)
\right\Vert ^{2}\right)  ^{1/2}\nonumber\\
&  \leq\frac{2}{n}\left(  \sum_{g=1}^{G}E\left\Vert \widetilde{Y}%
_{g}\widetilde{Y}_{g}^{\prime}\right\Vert ^{2}\right)  ^{1/2}.\label{evv3}%
\end{align}
Using the bounds $\left\Vert \widetilde{Y}_{g}\widetilde{Y}_{g}^{\prime
}\right\Vert \leq n\varepsilon$ and$\left\Vert \widetilde{Y}_{g}%
\widetilde{Y}_{g}^{\prime}\right\Vert \leq\left\Vert \widetilde{X}_{g}^{\ast
}\right\Vert ^{2}$, we deduce $\left\Vert \widetilde{Y}_{g}\widetilde{Y}%
_{g}^{\prime}\right\Vert ^{2}\leq n\varepsilon\left\Vert \widetilde{X}%
_{g}^{\ast}\right\Vert ^{2}$. Thus (\ref{evv3}) is bounded by%
\begin{align*}
2\varepsilon^{1/2}\left(  \frac{1}{n}\sum_{g=1}^{G}E\left\Vert \widetilde{X}%
_{g}^{\ast}\right\Vert ^{2}\right)  ^{1/2} &  =2\varepsilon^{1/2}\left(
\frac{1}{n}E\left\Vert \sum_{g=1}^{G}\widetilde{X}_{g}^{\ast}\right\Vert
^{2}\right)  ^{1/2}\\
&  =2\varepsilon^{1/2}\left(  n\text{var}\left(  \overline{X}_{n}^{\ast
}\right)  \right)  ^{1/2}\\
&  =2\varepsilon^{1/2}\left(  \text{tr}I_{p}\right)  ^{1/2}\\
&  =\delta
\end{align*}
The first equality holds because $\widetilde{X}_{g}^{\ast}$ are independent
and mean zero, and the second and third use the definition of $\overline
{X}_{n}^{\ast}$. The final equality is $\varepsilon=\delta^{2}/4p$.

Together, we have shown that for $n$ sufficiently large,%
\[
E\left\Vert \widetilde{\Omega}_{n}^{\ast}-I_{p}\right\Vert \leq3\delta
\]
and hence (\ref{v2}) by Markov's Inequality.

By the continuous mapping theorem%
\[
\widetilde{\Omega}_{n}^{\ast-1/2}\xrightarrow{p}I_{p}^{-1/2}=I_{p}%
\]
and%
\[
\left\Vert \Omega_{n}^{-1/4}\widetilde{\Omega}_{n}^{\ast-1/2}\Omega_{n}%
^{1/4}-I_{p}\right\Vert =\left\Vert \widetilde{\Omega}_{n}^{\ast-1/2}%
-I_{p}\right\Vert \xrightarrow{p}0.
\]
Combined with Theorem \ref{clt} we find%
\begin{align*}
&  \widetilde{\Omega}_{n}^{-1/2}\sqrt{n}\overline{X}_{n}\\
&  =\widetilde{\Omega}_{n}^{-1/2}\Omega_{n}^{1/2}\Omega_{n}^{-1/2}\sqrt
{n}\overline{X}_{n}\\
&  =\Omega_{n}^{-1/4}\widetilde{\Omega}_{n}^{\ast-1/2}\Omega_{n}^{1/4}%
\Omega_{n}^{-1/2}\sqrt{n}\overline{X}_{n}\\
&  \xrightarrow {d}N\left(  \mathbf{0},I_{p}\right)
\end{align*}
This is (\ref{v3}).\qquad$\blacksquare$

\bigskip

\noindent\textbf{Proof of Theorem \ref{cov2}:} Since the estimator
$\widehat{\Omega}_{n}$ is invariant to $\mu$, without loss of generality we
assume $\mu=0$. In this case%
\[
\widehat{\Omega}_{n}=\widetilde{\Omega}_{n}-\frac{1}{n}\sum_{g=1}^{G}n_{g}%
^{2}\overline{X}_{n}\overline{X}_{n}^{\prime}.
\]
Then by the triangle inequality, Theorem \ref{cov}, Theorem \ref{clt}, and
(\ref{ng2n}),%
\begin{align*}
&  \left\Vert \Omega_{n}^{-1/2}\widehat{\Omega}_{n}\Omega_{n}^{-1/2}%
-I_{p}\right\Vert \\
&  \leq\left\Vert \Omega_{n}^{-1/2}\widetilde{\Omega}_{n}\Omega_{n}%
^{-1/2}-I_{p}\right\Vert \\
&  +\left(  \frac{1}{n^{2}}\sum_{g=1}^{G}n_{g}^{2}\right)  \left\Vert
\Omega_{n}^{-1/2}\sqrt{n}\overline{X}_{n}\right\Vert ^{2}\\
&  \leq o_{p}(1).
\end{align*}
This is (\ref{v4}). Equation (\ref{v5}) follows as in the proof of
(\ref{v3}).\qquad$\blacksquare$

\bigskip

\noindent\textbf{Proof of Theorem \ref{ulln}:} Define the cluster sums
$\widetilde{f}_{g}(\theta)=\sum_{i=1}^{n_{g}}$ $f(X_{gi},\theta)$ so that
$\overline{f}_{n}(\theta)=\frac{1}{n}\sum_{g=1}^{G}\widetilde{f}_{g}(\theta) $
where $\widetilde{f}_{g}(\theta)$ are mutually independent.

Andrews (1992, Theorem 3) shows that (\ref{supu}) holds if $\Theta$ is totally
bounded,
\[
\left\Vert \frac{1}{n}\sum_{g=1}^{G}\left(  \widetilde{f}_{g}(\theta
)-E\widetilde{f}_{g}(\theta)\right)  \right\Vert \rightarrow_{p}0
\]
and for all $\theta_{1},\theta_{2}\in\Theta\,$,
\begin{equation}
\left\Vert \widetilde{f}_{g}(\theta_{1})-\widetilde{f}_{g}(\theta
_{2})\right\Vert \leq A_{g}h\left(  \left\Vert \theta_{1}-\theta
_{2}\right\Vert \right)  \label{uu}%
\end{equation}
where $h(u)\downarrow0$ as $u\downarrow0$ and $\frac{1}{n}\sum_{g=1}%
^{G}E\left(  A_{g}\right)  \leq A<\infty$. The total boundedness condition
holds by assumption and the WLLN holds by\ Theorem \ref{wlln} under Assumption
\ref{A1} and (\ref{f1}), so it only remains to establish the Lipschitz
condition (\ref{uu}). Indeed, using the triangle inequality and (\ref{lip1})%
\begin{align*}
\left\Vert \widetilde{f}_{g}(\theta_{2})-\widetilde{f}_{g}(\theta
_{1})\right\Vert  &  =\left\Vert \sum_{j=1}^{n_{g}}\left(  f(X_{gj},\theta
_{2})-f(X_{gj},\theta_{1})\right)  \right\Vert \\
&  \leq\sum_{j=1}^{n_{g}}\left\Vert f(X_{gj},\theta_{2})-f(X_{gj},\theta
_{1})\right\Vert \\
&  \leq\sum_{j=1}^{n_{g}}A(X_{gj})h\left(  \left\Vert \theta_{1}-\theta
_{2}\right\Vert \right) \\
&  =A_{g}h\left(  \left\Vert \theta_{1}-\theta_{2}\right\Vert \right)
\end{align*}
where $A_{g}=\sum_{j=1}^{n_{g}}A(X_{gj})$. Notice that%
\[
\frac{1}{n}\sum_{g=1}^{G}E\left(  A_{g}\right)  =\frac{1}{n}\sum_{g=1}^{G}%
\sum_{j=1}^{n_{g}}EA(X_{gj}) \leq C
\]
since $\sup_{i}EA(X_{i})\leq C$. This verifies (\ref{uu}) and hence
(\ref{supu}) holds.\qquad$\blacksquare$

\bigskip

\noindent\textbf{Proof of Theorem \ref{ullnv}:} Without loss of generality,
assume $\mu(\theta)=0$.

We first examine the case with no estimated mean (\ref{vu3}). Andrews (1992,
Theorem 3) shows that (\ref{vu3}) holds if for all $\theta\in\Theta$%
\begin{equation}
\left\Vert \widetilde{\Omega}_{n}(\theta)-E\widetilde{\Omega}_{n}%
(\theta)\right\Vert \rightarrow_{p}0, \label{rfr}%
\end{equation}
and for all $\theta_{1}$, $\theta_{2}\in\Theta$,
\begin{equation}
\left\Vert \widetilde{f}_{g}(\theta_{1})\widetilde{f}_{g}(\theta_{1})^{\prime
}-\widetilde{f}_{g}(\theta_{2})\widetilde{f}(\theta_{2})^{\prime}\right\Vert
\leq A_{g}h(\left\Vert \theta_{1}-\theta_{2}\right\Vert ) \label{ss}%
\end{equation}
with $h(u)\downarrow0$ as $u\downarrow0$ and $\frac{1}{n}\sum_{g=1}^{G}%
EA_{g}\leq A<\infty$. We now establish (\ref{rfr}) and (\ref{ss}).

Take (\ref{rfr}). Fix $\theta\in\Theta$. For brevity, suppress the dependence
of $\widetilde{f}_{g}(\theta)$ on $\theta$. Fix $\delta>0$. Set $\varepsilon
=\left(  \delta/C\right)  ^{2}$. Define $\widetilde{h}_{g}=\widetilde{f}%
_{g}1\left(  \left\Vert \widetilde{f}_{g}\right\Vert \leq\sqrt{n\varepsilon
}\right)  $. Then
\begin{equation}
\widetilde{\Omega}_{n}(\theta)=\frac{1}{n}\sum_{g=1}^{G}\widetilde{h}%
_{g}\widetilde{h}_{g}^{\prime}+\frac{1}{n}\sum_{g=1}^{G}\widetilde{f}%
_{g}\widetilde{f}_{g}^{\prime}1\left(  \left\Vert \widetilde{f}_{g}\right\Vert
>\sqrt{n\varepsilon}\right)  .\nonumber
\end{equation}
By the triangle inequality%
\begin{align}
E\left\Vert \widetilde{\Omega}_{n}(\theta)-E\widetilde{\Omega}_{n}%
(\theta)\right\Vert  &  =\frac{1}{n}E\left\Vert \sum_{g=1}^{G}\left(
\widetilde{h}_{g}\widetilde{h}_{g}^{\prime}-E\widetilde{h}_{g}\widetilde{h}%
_{g}^{\prime}\right)  \right\Vert \label{omi1}\\
&  +\frac{2}{n}\sum_{g=1}^{G}E\left(  \left\Vert \widetilde{f}_{g}\right\Vert
^{2}1\left(  \left\Vert \widetilde{f}_{g}\right\Vert >\sqrt{n\varepsilon
}\right)  \right)  . \label{omi2}%
\end{align}
Take (\ref{omi1}). Assumption (\ref{fr}) and the $C_{r}$ inequality allow us
to deduce that
\begin{equation}
E\left\Vert \widetilde{f}_{g}\right\Vert ^{2}\leq Cn_{g}^{2} \label{efr}%
\end{equation}
for some $C<\infty$. Using Jensen's inequality, the assumption the clusters
are independent and thus uncorrelated, the bounds $\left\Vert \widetilde{h}%
_{g}\right\Vert \leq\sqrt{n\varepsilon}$ and $\left\Vert \widetilde{h}%
_{g}\right\Vert \leq\left\Vert \widetilde{f}_{g}\right\Vert $, (\ref{efr}),
(\ref{nbound}) with $r=2$ and the definition of $\varepsilon$, we obtain that
(\ref{omi1}) is bounded by%
\begin{align*}
\frac{1}{n}\left(  E\left\Vert \sum_{g=1}^{G}\left(  \widetilde{h}%
_{g}\widetilde{h}_{g}^{\prime}-E\widetilde{h}_{g}\widetilde{h}_{g}^{\prime
}\right)  \right\Vert ^{2}\right)  ^{1/2}  &  \leq\frac{1}{n}\left(
\sum_{g=1}^{G}E\left\Vert \widetilde{h}_{g}\right\Vert ^{4}\right)  ^{1/2}\\
&  \leq\varepsilon^{1/2}C^{1/2}\left(  \frac{1}{n}\sum_{g=1}^{G}n_{g}%
^{2}\right)  ^{1/2}\leq\delta.
\end{align*}
Take (\ref{omi2}). Lemma 1 implies that $\left\Vert n_{g}^{-1}\widetilde{f}%
_{g}\right\Vert ^{2}$ is uniformly integrable given Assumption (\ref{fr}).
This means we can pick $B$ sufficiently large so that
\begin{equation}
\sup_{g}E\left(  \left\Vert n_{g}^{-1}\widetilde{f}_{g}\right\Vert
^{2}1\left(  \left\Vert n_{g}^{-1}\widetilde{f}_{g}\right\Vert >B\right)
\right)  \leq\frac{\delta}{C} \label{uib}%
\end{equation}
Pick $n$ large enough so that
\[
\max_{g\leq G}\frac{n_{g}}{n^{1/2}}\leq\max_{g\leq G}\frac{n_{g}^{2}}{n^{1/2}%
}\leq\frac{\sqrt{\varepsilon}}{B}%
\]
which is feasible by (\ref{ng2n}). Then (\ref{omi2}) is bounded by%
\[
\frac{2}{n}\sum_{g=1}^{G}E\left(  \left\Vert \widetilde{f}_{g}\right\Vert
^{2}1\left(  \left\Vert n_{g}^{-1}\widetilde{f}_{g}\right\Vert >B\right)
\right)  \leq\frac{2}{n}\sum_{g=1}^{G}n_{g}^{2}\frac{\delta}{C}\leq2\delta,
\]
using (\ref{uib}) and (\ref{nbound}) with $r=2$. We have shown that
$E\left\Vert \widetilde{\Omega}_{n}(\theta)-E\widetilde{\Omega}_{n}%
(\theta)\right\Vert \leq3\delta$. Since $\delta$ is arbitrary, by Markov's
inequality, (\ref{rfr}) is shown.

Take (\ref{ss}). Fix any $\theta_{1},\theta_{2}\in\Theta$. Set $\widetilde{f}%
_{g}=\sup_{\theta\in\Theta}\left\Vert \widetilde{f}_{g}(\theta)\right\Vert $.
Using the triangle inequality and Assumption (\ref{lip1})%
\[
\left\Vert \widetilde{f}_{g}(\theta_{2})-\widetilde{f}_{g}(\theta
_{1})\right\Vert \leq\sum_{j=1}^{n_{g}}A(X_{gj})h\left(  \left\Vert \theta
_{1}-\theta_{2}\right\Vert \right)  .
\]
Then%
\begin{align*}
\left\Vert \widetilde{f}_{g}(\theta_{1})\widetilde{f}_{g}(\theta_{1})^{\prime
}-\widetilde{f}_{g}(\theta_{2})\widetilde{f}(\theta_{2})^{\prime}\right\Vert
&  \leq2\widetilde{f}_{g}\left\Vert \widetilde{f}_{g}(\theta_{2}%
)-\widetilde{f}(\theta_{1})\right\Vert \\
&  \leq2\widetilde{f}_{g}\left(  \sum_{j=1}^{n_{g}}A(X_{gj})\right)  h\left(
\left\Vert \theta_{1}-\theta_{2}\right\Vert \right)  .
\end{align*}
Hence (\ref{ss}) holds with $A_{g}=2\widetilde{f}_{g}\left(  \sum_{j=1}%
^{n_{g}}A(X_{gj})\right)  $.

It remains to show that $\frac{1}{n}\sum_{g=1}^{G}EA_{g}\leq A<\infty$.
Assumption (\ref{fr}) and the $C_{r}$ inequality allows us to deduce that
$E\widetilde{f}_{g}^{2}\leq Cn_{g}^{2}$. Applying Holder's inequality%
\[
EA_{g}\leq2\sum_{j=1}^{n_{g}}\left(  E\widetilde{f}_{g}^{2}\right)
^{1/2}\left(  EA^{2}(X_{gj})\right)  ^{1/2}\leq2Cn_{g}^{2}.
\]
Hence%
\[
\frac{1}{n}\sum_{g=1}^{G}EA_{g}\leq2C\frac{1}{n}\sum_{g=1}^{G}n_{g}^{2}%
\leq2C^{2}%
\]
by Assumption (\ref{nbound}) with $r=2$. This establishes (\ref{ss}).

By showing (\ref{rfr}) and (\ref{ss}) we have established (\ref{vu3}).

The case with estimated mean (\ref{vu2}) immediately follows from (\ref{vu3})
and Theorem \ref{ulln}. \qquad$\blacksquare$

\bigskip

\noindent\textbf{Proof of Theorem \ref{cltv}:} Define
\begin{align*}
\widetilde{f}_{g}^{\ast}  &  =\Omega_{n}^{-1/2}\widetilde{f}_{g}\\
\overline{f}_{G}^{\ast}  &  =\frac{1}{n}\sum_{g=1}^{G}\widetilde{f}_{g}^{\ast
}.
\end{align*}
Then%
\[
\Omega_{n}^{-1/2}\sqrt{n}\left(  \overline{f}_{G}-E\overline{f}_{G}\right)
=\sqrt{n}\left(  \overline{f}_{G}^{\ast}-E\overline{f}_{G}^{\ast}\right)
\]
where $n\text{var}\left(  \overline{f}_{G}^{\ast}\right)  =I_{p}$.

Since $\widetilde{f}_{g}^{\ast}$ are independent but not identically
distributed, we apply the multivariate Lindeberg-Feller central limit theorem
(e.g. Hansen (2018) Theorem 6.15). Since $\text{var}\left(  \sqrt{n}%
\overline{f}_{G}^{\ast}\right)  =I_{p}$ a sufficient condition for the CLT
(\ref{clt2}) is that for all $\varepsilon>0$%
\begin{align}
&  \frac{1}{n}\sum_{g=1}^{G}E\left(  \left\Vert \widetilde{f}_{g}^{\ast
}-E\widetilde{f}_{g}^{\ast}\right\Vert ^{2}1\left(  \left\Vert \widetilde{f}%
_{g}^{\ast}-E\widetilde{f}_{g}^{\ast}\right\Vert ^{2}\geq n\varepsilon\right)
\right) \nonumber\\
&  \leq\frac{1}{n\lambda}\sum_{g=1}^{G}E\left(  \left\Vert \widetilde{f}%
_{g}-E\widetilde{f}_{g}\right\Vert ^{2}1\left(  \left\Vert \widetilde{f}%
_{g}-E\widetilde{f}_{g}\right\Vert ^{2}\geq n\varepsilon\lambda\right)
\right)  \rightarrow0 \label{lind2}%
\end{align}
as $n\rightarrow\infty$.

Fix $\varepsilon>0$ and $\delta>0$. Pick $B$ sufficiently large so that%
\begin{equation}
\sup_{g}E\left(  \left\Vert n_{g}^{-2}\left(  \widetilde{f}_{g}-E\widetilde{f}%
_{g}\right)  \right\Vert ^{r}1\left(  \left\Vert n_{g}^{-2}\left(
\widetilde{f}_{g}-E\widetilde{f}_{g}\right)  \right\Vert >B\right)  \right)
\leq\frac{\delta\varepsilon^{r/2-1}\lambda^{r/2}}{C^{r/2}}. \label{cltbv}%
\end{equation}
which is feasible by Lemma \ref{uitheorem2} under (\ref{uirv}). Pick $n$ large
enough so that%
\begin{equation}
\max_{g\leq G}\frac{n_{g}^{2}}{n^{1/2}}\leq\frac{(\varepsilon\lambda)^{1/2}%
}{B} \label{cltb2v}%
\end{equation}
which is feasible by (\ref{ng4n}). Thus%
\begin{align}
&  \frac{1}{n\lambda}\sum_{g=1}^{G}E\left(  \left\Vert \widetilde{f}%
_{g}-E\widetilde{f}_{g}\right\Vert ^{2}1\left(  \left\Vert \widetilde{f}%
_{g}-E\widetilde{f}_{g}\right\Vert ^{2}\geq n\varepsilon\lambda\right)
\right) \label{lindv3}\\
&  =\frac{1}{n\lambda}\sum_{g=1}^{G}E\left(  \frac{\left\Vert \widetilde{f}%
_{g}-E\widetilde{f}_{g}\right\Vert ^{r}}{\left\Vert \widetilde{f}%
_{g}-E\widetilde{f}_{g}\right\Vert ^{r-2}}1\left(  \left\Vert \widetilde{f}%
_{g}-E\widetilde{f}_{g}\right\Vert \geq(n\varepsilon\lambda)^{1/2}\right)
\right) \nonumber\\
&  \leq\frac{1}{\varepsilon^{r/2-1}n^{r/2}\lambda^{r/2}}\sum_{g=1}^{G}E\left(
\left\Vert \widetilde{f}_{g}-E\widetilde{f}_{g}\right\Vert ^{r}1\left(
\left\Vert \widetilde{f}_{g}-E\widetilde{f}_{g}\right\Vert \geq\left(
n\varepsilon\lambda\right)  ^{1/2}\right)  \right) \nonumber\\
&  \leq\frac{1}{\varepsilon^{r/2-1}n^{r/2}\lambda^{r/2}}\sum_{g=1}^{G}%
n_{g}^{2r}E\left(  \left\Vert n_{g}^{-2}\left(  \widetilde{f}_{g}%
-E\widetilde{f}_{g}\right)  \right\Vert ^{r}1\left(  \left\Vert n_{g}%
^{-2}\left(  \widetilde{f}_{g}-E\widetilde{f}_{g}\right)  \right\Vert \geq
B\right)  \right) \nonumber\\
&  \leq\frac{\delta}{C^{r/2}}\frac{\sum_{g=1}^{G}n_{g}^{2r}}{n^{r/2}%
}\nonumber\\
&  \leq\delta.\nonumber
\end{align}
The second inequality is (\ref{cltb2v}), the third is (\ref{cltbv}), and the
final is\ (\ref{nboundv}). Since $\varepsilon$ and $\delta$ are arbitrary we
have established (\ref{lind2}) and hence (\ref{clt2}).\qquad$\blacksquare$

\bigskip

The proofs of Theorems 8-13 are presented in the Supplemental Appendix.

\bigskip\newpage%

\setcounter{page}{1}

\begin{center}
	{\Large {\textbf{Supplemental Appendix}} }
	
	{\Large \bigskip}
	
	{\Large {\textbf{Asymptotic Theory for Clustered Samples}} }
\end{center}

\bigskip

\begin{center}
	$%
	\begin{array}
	[c]{ccc}%
	\text{Bruce E. Hansen} & \hspace{0.2in} & \text{Seojeong Lee}\\
	\text{University of Wisconsin} & \hspace{0.2in} & \text{University of New
		South Wales}%
	\end{array}
	$ \medskip
\end{center}

In this supplemental appendix we present proofs of Theorems 8-13.

\bigskip

\noindent\noindent\textbf{Proof of Theorem 8:} Write
\[
\widehat{\boldsymbol{\beta}}-\boldsymbol{\beta}=\left(  \widehat{Q}%
_{n}^{\prime}\widehat{W}_{n}^{-1}\widehat{Q}_{n}\right)  ^{-1}\widehat{Q}%
_{n}^{\prime}\widehat{W}_{n}^{-1}\widehat{S}_{n}%
\]
where
\[
\widehat{S}_{n}=\frac{1}{n}\sum_{g=1}^{G}\boldsymbol{Z}_{g}^{\prime
}\boldsymbol{e}_{g}=\frac{1}{n}\sum_{i=1}^{n}\boldsymbol{z}_{i}e_{i}.
\]

The random variables $(\boldsymbol{z}_{i}\boldsymbol{x}_{i}^{\prime
},\boldsymbol{z}_{i}\boldsymbol{z}_{i}^{\prime},\boldsymbol{z}_{i}e_{i})$ are
uniformly integrable under the assumptions. By Theorem 1
\begin{align}
\left\Vert \widehat{S}_{n}\right\Vert  &  \xrightarrow{p}0,\label{2slscw}\\
\left\Vert \widehat{Q}_{n}-Q_{n}\right\Vert  &
\xrightarrow{p}0,\label{2slscq}\\
\left\Vert \widehat{W}_{n}-W_{n}\right\Vert  &  \xrightarrow{p}0.
\label{2slsw}%
\end{align}

We first show
\begin{equation}
\left\Vert \widehat{W}_{n}^{-1}-W_{n}^{-1}\right\Vert \xrightarrow{p}0.
\label{2slscw1}%
\end{equation}
By $\lambda_{\min}(W_{n})\geq C$ and (\ref{2slsw}),
\[
\left\Vert W_{n}^{-1/2}\widehat{W}_{n}W_{n}^{-1/2}-I_{l}\right\Vert
\leq\left\Vert W_{n}^{-1}\right\Vert \left\Vert \widehat{W}_{n}-W_{n}%
\right\Vert \leq C^{-1}\left\Vert \widehat{W}_{n}-W_{n}\right\Vert
\xrightarrow{p}0.
\]
By the continuous mapping theorem,
\begin{equation}
\left(  W_{n}^{-1/2}\widehat{W}_{n}W_{n}^{-1/2}\right)  ^{-1}%
\xrightarrow{p}I_{l}^{-1}=I_{l}.\nonumber
\end{equation}
Thus,
\begin{align*}
\left\Vert \widehat{W}_{n}^{-1}-W_{n}^{-1}\right\Vert  &  =\left\Vert
W_{n}^{-1/2}\left(  W_{n}^{1/2}\widehat{W}_{n}^{-1}W_{n}^{1/2}-I_{l}\right)
W_{n}^{-1/2}\right\Vert \\
&  \leq C^{-1}\left\Vert W_{n}^{1/2}\widehat{W}_{n}^{-1}W_{n}^{1/2}%
-I_{l}\right\Vert \xrightarrow{p}0.
\end{align*}

Next we show
\begin{equation}
(Q_{n}^{\prime}W_{n}^{-1}Q_{n})^{1/2}(\widehat{Q}_{n}^{\prime}\widehat{W}%
_{n}^{-1}\widehat{Q}_{n})^{-1}(Q_{n}^{\prime}W_{n}^{-1}Q_{n})^{1/2}%
\xrightarrow{p}I_{k}. \label{2slscqwq1}%
\end{equation}
By the continuous mapping theorem, (\ref{2slscqwq1}) is equivalent to show
\begin{equation}
(Q_{n}^{\prime}W_{n}^{-1}Q_{n})^{-1/2}(\widehat{Q}_{n}^{\prime}\widehat{W}%
_{n}^{-1}\widehat{Q}_{n})(Q_{n}^{\prime}W_{n}^{-1}Q_{n})^{-1/2}%
\xrightarrow{p}I_{k}. \label{2slscqwq2}%
\end{equation}
Under $\lambda_{\min}(W_{n})\geq C>0$ and the full column rank condition,
\begin{equation}
\lambda_{\min}\left(  Q_{n}^{\prime}W_{n}^{-1}Q_{n}\right)  \geq C>0.
\label{2slscqwq3}%
\end{equation}
Since
\begin{align}
&  \left\Vert (Q_{n}^{\prime}W_{n}^{-1}Q_{n})^{-1/2}(\widehat{Q}_{n}^{\prime
}\widehat{W}_{n}^{-1}\widehat{Q}_{n})(Q_{n}^{\prime}W_{n}^{-1}Q_{n}%
)^{-1/2}-I_{k}\right\Vert \nonumber\\
&  \leq C^{-1}\left\Vert \left(  \widehat{Q}_{n}^{\prime}\widehat{W}_{n}%
^{-1}\widehat{Q}_{n}-Q_{n}^{\prime}W_{n}^{-1}Q_{n}\right)  \right\Vert
\nonumber
\end{align}
By (\ref{2slscqwq3}) it is sufficient for (\ref{2slscqwq2}) to show
\begin{equation}
\widehat{Q}_{n}^{\prime}\widehat{W}_{n}^{-1}\widehat{Q}_{n}-Q_{n}^{\prime
}W_{n}^{-1}Q_{n}\xrightarrow{p}0. \label{2slscqwq4}%
\end{equation}
Observe that by Cauchy-Schwarz inequality,
\begin{align}
\left\Vert Q_{n}\right\Vert  &  \leq\frac{1}{n}\sum_{g=1}^{G}\sum_{j=1}%
^{n_{g}}E\left\Vert \boldsymbol{z}_{gj}\boldsymbol{x}_{gj}^{\prime}\right\Vert
\nonumber\\
&  \leq\sup_{i}E\left\Vert \boldsymbol{z}_{i}\boldsymbol{x}_{i}^{\prime
}\right\Vert \nonumber\\
&  \leq\sup_{i}\left(  E\left\Vert \boldsymbol{z}_{i}\right\Vert ^{2}\right)
^{1/2}\left(  E\left\Vert \boldsymbol{x}_{i}\right\Vert ^{2}\right)
^{1/2}<\infty. \label{Qbd}%
\end{align}
By centering $\widehat{Q}_{n}$ and $\widehat{W}_{n}^{-1}$ around $Q_{n}$ and
$W_{n}^{-1}$,
\begin{align*}
&  \left\Vert \widehat{Q}_{n}^{\prime}\widehat{W}_{n}^{-1}\widehat{Q}%
_{n}-Q_{n}^{\prime}W_{n}^{-1}Q_{n}\right\Vert \\
&  \leq\left\Vert \left(  \widehat{Q}_{n}-Q_{n}\right)  ^{\prime}\left(
\widehat{W}_{n}^{-1}-W_{n}^{-1}\right)  \left(  \widehat{Q}_{n}-Q_{n}\right)
\right\Vert \\
&  +2\left\Vert \left(  \widehat{Q}_{n}-Q_{n}\right)  ^{\prime}\left(
\widehat{W}_{n}^{-1}-W_{n}^{-1}\right)  Q_{n}\right\Vert +\left\Vert \left(
\widehat{Q}_{n}-Q_{n}\right)  ^{\prime}W_{n}^{-1}\left(  \widehat{Q}_{n}%
-Q_{n}\right)  \right\Vert \\
&  +\left\Vert \left(  \widehat{Q}_{n}-Q_{n}\right)  ^{\prime}W_{n}^{-1}%
Q_{n}\right\Vert +\left\Vert Q_{n}^{\prime}W_{n}^{-1}\left(  \widehat{Q}%
_{n}-Q_{n}\right)  \right\Vert +\left\Vert Q_{n}^{\prime}\left(
\widehat{W}_{n}^{-1}-W_{n}^{-1}\right)  Q_{n}\right\Vert \\
&  \leq o_{p}(1)
\end{align*}
by (\ref{2slscq}), (\ref{2slscw1}), $\Vert W_{n}^{-1}\Vert\leq C^{-1}$, and
\eqref{Qbd}. This is (\ref{2slscqwq4}).

Lastly, we show
\begin{equation}
\left\Vert \widehat{Q}_{n}^{\prime}\widehat{W}_{n}^{-1}\widehat{S}%
_{n}\right\Vert \xrightarrow{p}0. \label{2slscqws}%
\end{equation}
Using (\ref{2slscw}), (\ref{2slscq}), (\ref{2slscw1}), and $\Vert W_{n}%
^{-1}\Vert\leq C^{-1}$,
\begin{align*}
\left\Vert \widehat{Q}_{n}^{\prime}\widehat{W}_{n}^{-1}\widehat{S}%
_{n}\right\Vert  &  \leq\left\Vert Q_{n}^{\prime}W_{n}^{-1}\widehat{S}%
_{n}\right\Vert \\
&  +\left\Vert Q_{n}^{\prime}(\widehat{W}_{n}^{-1}-W_{n}^{-1})\widehat{S}%
_{n}+(\widehat{Q}_{n}-Q_{n})^{\prime}W_{n}^{-1}\widehat{S}_{n}+(\widehat{Q}%
_{n}-Q_{n})^{\prime}\left(  \widehat{W}_{n}^{-1}-W_{n}^{-1}\right)
\widehat{S}_{n}\right\Vert \\
&  \leq\left(  O(1)C^{-1}+o_{p}(1)\right)  \left\Vert \widehat{S}%
_{n}\right\Vert \xrightarrow{p}0
\end{align*}
as required.

Combining the results (\ref{2slscqwq1}), (\ref{2slscqwq3}), and
(\ref{2slscqws}),
\begin{align*}
\left\Vert \widehat{\boldsymbol{\beta}}-\boldsymbol{\beta}\right\Vert  &
\leq\left(  \left\Vert \left(  Q_{n}^{\prime}W_{n}^{-1}Q_{n}\right)  ^{-1}
\right\Vert + \left\Vert \left(  Q_{n}^{\prime}W_{n}^{-1}Q_{n}\right)
^{1/2}\left(  \widehat{Q}_{n}^{\prime}\widehat{W}_{n}^{-1}\widehat{Q}%
_{n}\right)  ^{-1}\left(  Q_{n}^{\prime}W_{n}^{-1}Q_{n}\right)  ^{1/2}%
-I_{k}\right\Vert \right) \\
&  \cdot\left\Vert \widehat{Q}_{n}^{\prime}\widehat{W}_{n}^{-1}\widehat{S}%
_{n}\right\Vert \\
&  \leq\left(  C^{-1} + o_{p}(1)\right)  o_{p}(1).
\end{align*}
This completes the proof. \qquad$\blacksquare$

\bigskip

\noindent\textbf{Proof of Theorem 9:} We start by showing some
useful results. Since $Q_{n}^{\prime}W_{n}^{-1}Q_{n} = O(1)$, by
(\ref{2slscqwq4}),
\begin{align}
\label{qwq}\left(  \widehat{Q}_{n}^{\prime}\widehat{W}_{n}^{-1}\widehat{Q}%
_{n}\right)  ^{-1}  &  = \left(  Q_{n}^{\prime}W_{n}^{-1}Q_{n}\right)
^{-1}\left(  I_{k} + Q_{n}^{\prime}W_{n}^{-1}Q_{n}\left(  (\widehat{Q}%
_{n}^{\prime}\widehat{W}_{n}^{-1}\widehat{Q}_{n})^{-1} - (Q_{n}^{\prime}%
W_{n}^{-1}Q_{n})^{-1}\right)  \right) \nonumber\\
&  = \left(  Q_{n}^{\prime}W_{n}^{-1}Q_{n}\right)  ^{-1}\left(  I_{k} +
o_{p}(1)\right)  .
\end{align}
In addition, by (\ref{2slscq}) and (\ref{2slscw1})
\begin{align}
\label{qw}\widehat{Q}_{n}^{\prime}\widehat{W}_{n}^{-1}  &  = Q_{n}^{\prime
}W_{n}^{-1} + Q_{n}^{\prime}W_{n}^{-1}Q_{n}\left(  Q_{n}^{\prime}W_{n}%
^{-1}Q_{n}\right)  ^{-1}\left(  \widehat{Q}_{n}^{\prime}\widehat{W}_{n}%
^{-1}-Q_{n}^{\prime}W_{n}^{-1}\right) \nonumber\\
&  =Q_{n}^{\prime}W_{n}^{-1}\left(  I_{l} + o_{p}(1)\right)  .
\end{align}

By \eqref{qwq} and \eqref{qw},
\begin{align}
&  \left(  \widehat{Q}_{n}^{\prime}\widehat{W}_{n}^{-1}\widehat{Q}_{n}\right)
^{-1}\widehat{Q}_{n}^{\prime}\widehat{W}_{n}^{-1}\sqrt{n}\widehat{S}%
_{n}\nonumber\\
&  =\left(  Q_{n}^{\prime}W_{n}^{-1}Q_{n}\right)  ^{-1}Q_{n}^{\prime}%
W_{n}^{-1}\sqrt{n}\widehat{S}_{n}\label{2slse1}\\
&  +\left(  Q_{n}^{\prime}W_{n}^{-1}Q_{n}\right)  ^{-1}o_{p}(1)Q_{n}^{\prime
}W_{n}^{-1}\sqrt{n}\widehat{S}_{n}+\left(  Q_{n}^{\prime}W_{n}^{-1}%
Q_{n}\right)  ^{-1}Q_{n}^{\prime}W_{n}^{-1}o_{p}(1)\sqrt{n}\widehat{S}%
_{n}\label{2slse2}\\
&  +\left(  Q_{n}^{\prime}W_{n}^{-1}Q_{n}\right)  ^{-1}o_{p}(1)Q_{n}^{\prime
}W_{n}^{-1}o_{p}(1)\sqrt{n}\widehat{S}_{n} \label{2slse3}%
\end{align}
Let
\[
R_{n}^{\ast}=V_{n}^{1/2}R_{n}\left(  R_{n}^{\prime}V_{n}R_{n}\right)
^{-1/2}.
\]
First take (\ref{2slse1}). Since
\[
n\text{var}\left(  V_{n}^{-1/2}\left(  Q_{n}^{\prime}W_{n}^{-1}Q_{n}\right)
^{-1}Q_{n}^{\prime}W_{n}^{-1}\widehat{S}_{n}\right)  =I_{k},
\]
we apply Theorem 2 to find
\begin{equation}
R_{n}^{\ast\prime}V_{n}^{-1/2}\left(  Q_{n}^{\prime}W_{n}^{-1}Q_{n}\right)
^{-1}Q_{n}^{\prime}W_{n}^{-1}\sqrt{n}\widehat{S}_{n}\xrightarrow{d}N(0,I_{q})
\label{2slse1r}%
\end{equation}
In addition, \eqref{2slse1r} implies that (\ref{2slse2}) and (\ref{2slse3})
are bounded by $O_{p}(1)o_{p}(1)=o_{p}(1)$. Thus (31) follows under
the assumptions.

For (32) we show that
\begin{equation}
\left\Vert V_{n}^{-1/2}\widehat{V}_{n}V_{n}^{-1/2}-I_{k}\right\Vert
\xrightarrow{p}0. \label{vvv}%
\end{equation}
We first show that
\begin{equation}
\left\Vert \Omega_{n}^{-1}\left(  \widehat{\Omega}_{n}-\Omega_{n}\right)
\right\Vert \xrightarrow{p}0 \label{omega1}%
\end{equation}
which is equivalent of showing
\[
\left\Vert \Omega_{n}^{-1/2}\widehat{\Omega}_{n}\Omega_{n}^{-1/2}%
-I_{l}\right\Vert \xrightarrow{p}0.
\]
Define
\[
\widetilde{\Omega}_{n}=\frac{1}{n}\sum_{g=1}^{G}\boldsymbol{Z}_{g}^{\prime
}\boldsymbol{e}_{g}\boldsymbol{e}_{g}^{\prime}\boldsymbol{Z}_{g}%
\]
and%
\begin{align*}
\widetilde{\Omega}_{n}^{\ast}  &  =\frac{1}{n}\sum_{g=1}^{G}\boldsymbol{Z}%
_{g}^{\prime}\left(  \widehat{\boldsymbol{e}}_{g}-\boldsymbol{e}_{g}\right)
\left(  \widehat{\boldsymbol{e}}_{g}-\boldsymbol{e}_{g}\right)  ^{\prime
}\boldsymbol{Z}_{g}\\
&  =\frac{1}{n}\sum_{g=1}^{G}\boldsymbol{Z}_{g}^{\prime}\boldsymbol{X}%
_{g}\left(  \widehat{\boldsymbol{\beta}}-\boldsymbol{\beta}\right)  \left(
\widehat{\boldsymbol{\beta}}-\boldsymbol{\beta}\right)  ^{\prime
}\boldsymbol{X}_{g}^{\prime}\boldsymbol{Z}_{g}.
\end{align*}
By the triangle and Cauchy-Schwarz inequalities%
\begin{align}
\left\Vert \Omega_{n}^{-1/2}\widehat{\Omega}_{n}\Omega_{n}^{-1/2}%
-I_{l}\right\Vert  &  \leq\left\Vert \Omega_{n}^{-1/2}\widetilde{\Omega}%
_{n}\Omega_{n}^{-1/2}-I_{l}\right\Vert +\left\Vert \Omega_{n}^{-1/2}\frac
{1}{n}\sum_{g=1}^{G}\boldsymbol{Z}_{g}^{\prime}\left(  \widehat{\boldsymbol{e}%
}_{g}\widehat{\boldsymbol{e}}_{g}^{\prime}-\boldsymbol{e}_{g}\boldsymbol{e}%
_{g}^{\prime}\right)  \boldsymbol{Z}_{g}\Omega_{n}^{-1/2}\right\Vert
\nonumber\\
&  \leq\left\Vert \Omega_{n}^{-1/2}\widetilde{\Omega}_{n}\Omega_{n}%
^{-1/2}-I_{l}\right\Vert +2\left\Vert \Omega_{n}^{-1/2}\widetilde{\Omega}%
_{n}\Omega_{n}^{-1/2}\right\Vert ^{1/2}\left\Vert \Omega_{n}^{-1/2}%
\widetilde{\Omega}_{n}^{\ast}\Omega_{n}^{-1/2}\right\Vert ^{1/2}\nonumber\\
&  +\left\Vert \Omega_{n}^{-1/2}\widetilde{\Omega}_{n}^{\ast}\Omega_{n}%
^{-1/2}\right\Vert .\nonumber
\end{align}
Under the assumption, Theorem 3 implies that
\[
\Omega_{n}^{-1/2}\widetilde{\Omega}_{n}\Omega_{n}^{-1/2}\xrightarrow{p}I_{l}.
\]
The proof of \eqref{omega1} is completed by showing that%
\[
\left\Vert \Omega_{n}^{-1/2}\widetilde{\Omega}_{n}^{\ast}\Omega_{n}%
^{-1/2}\right\Vert \xrightarrow{p}0.
\]
Since $\Vert(Q_{n}^{\prime}W_{n}^{-1}Q_{n})^{-1}\Vert\leq C^{-1}$ and $\Vert
W_{n}^{-1}\Vert\leq C^{-1}$,
\begin{align*}
\left\Vert \Omega_{n}^{-1/2}\widetilde{\Omega}_{n}^{\ast}\Omega_{n}%
^{-1/2}\right\Vert  &  =\left\Vert \Omega_{n}^{-1/2}\frac{1}{n}\sum_{g=1}%
^{G}\boldsymbol{Z}_{g}^{\prime}\boldsymbol{X}_{g}\left(
\widehat{\boldsymbol{\beta}}-\boldsymbol{\beta}\right)  \left(
\widehat{\boldsymbol{\beta}}-\boldsymbol{\ \beta}\right)  ^{\prime
}\boldsymbol{X}_{g}^{\prime}\boldsymbol{Z}_{g}\Omega_{n}^{-1/2}\right\Vert \\
&  \leq\left\Vert V_{n}^{-1/2}\sqrt{n}\left(  \widehat{\boldsymbol{\beta}%
}-\boldsymbol{\beta}\right)  \right\Vert ^{2}\left\Vert \frac{1}{n^{2}}%
\sum_{g=1}^{G}\boldsymbol{X}_{g}^{\prime}\boldsymbol{Z}_{g}\Omega_{n}%
^{-1}\boldsymbol{Z}_{g}^{\prime}\boldsymbol{X}_{g}V_{n}\right\Vert \\
&  \leq O_{p}(1)C^{-4}\frac{1}{n^{2}}\sum_{g=1}^{G}\left\Vert \boldsymbol{Z}%
_{g}^{\prime}\boldsymbol{X}_{g}\right\Vert ^{2}\\
&  =o_{p}(1)
\end{align*}
since by Minkowski's and Cauchy-Schwarz inequalities, and $E\left\Vert
\boldsymbol{x}_{gi}\right\Vert ^{4}\leq C$, and $E\left\Vert \boldsymbol{z}%
_{gi}\right\Vert ^{4}\leq C$,
\begin{align}
E\left\Vert \boldsymbol{Z}_{g}^{\prime}\boldsymbol{X}_{g}\right\Vert
^{2}=E\left\Vert \sum_{i=1}^{n_{g}}\boldsymbol{z}_{gi}\boldsymbol{x}%
_{gi}^{\prime}\right\Vert ^{2}  &  \leq\left(  \sum_{i=1}^{n_{g}}\left(
E\left\Vert \boldsymbol{z}_{gi}\boldsymbol{x}_{gi}^{\prime}\right\Vert
^{2}\right)  ^{1/2}\right)  ^{2}\nonumber\\
&  \leq\left(  \sum_{i=1}^{n_{g}}\left(  E\left\Vert \boldsymbol{z}%
_{gi}\right\Vert ^{2}\left\Vert \boldsymbol{x}_{gi}\right\Vert ^{2}\right)
^{1/2}\right)  ^{2}\nonumber\\
&  \leq\left(  \sum_{i=1}^{n_{g}}\left(  E\left\Vert \boldsymbol{z}%
_{gi}\right\Vert ^{4}\right)  ^{1/4}\left(  E\left\Vert \boldsymbol{x}%
_{gi}\right\Vert ^{4}\right)  ^{1/4}\right)  ^{2}\nonumber\\
&  \leq Cn_{g}^{2}\nonumber
\end{align}
so
\[
\frac{1}{n^{2}}\sum_{g=1}^{G}E\left\Vert \boldsymbol{Z}_{g}^{\prime
}\boldsymbol{X}_{g}\right\Vert ^{2}\leq\frac{C}{n^{2}}\sum_{g=1}^{G}n_{g}%
^{2}=o(1)
\]
by (6).

By \eqref{omega1},
\begin{align}
\label{omega0}\widehat{\Omega}_{n}  &  = \Omega_{n}\left(  I_{l} + \Omega
_{n}^{-1}\left(  \widehat{\Omega}_{n} - \Omega_{n}\right)  \right) \nonumber\\
&  = \Omega_{n}(I_{l} + o_{p}(1)).
\end{align}

By (\ref{qwq}), (\ref{qw}), \eqref{omega0}, and the triangle inequality
\begin{align*}
&  \left\Vert V_{n}^{-1/2}\widehat{V}_{n}V_{n}^{-1/2}-I_{k}\right\Vert \\
&  =\left\Vert V_{n}^{-1/2}\left(  \widehat{Q}_{n}^{\prime}\widehat{W}%
_{n}^{-1}\widehat{Q}_{n}\right)  ^{-1}\widehat{Q}_{n}^{\prime}\widehat{W}%
_{n}^{-1}\widehat{\Omega}_{n}\widehat{W}_{n}^{-1}\widehat{Q}_{n}\left(
\widehat{Q}_{n}^{\prime}\widehat{W}_{n}^{-1}\widehat{Q}_{n}\right)  ^{-1}%
V_{n}^{-1/2}-I_{k}\right\Vert \\
&  =\left\Vert V_{n}^{-1/2}\left(  Q_{n}^{\prime}W_{n}^{-1}Q_{n}\right)
^{-1}(I_{k}+o_{p}(1))Q_{n}^{\prime}W_{n}^{-1}(I_{l}+o_{p}(1))\Omega_{n}%
(I_{l}+o_{p}(1))\right. \\
&  \left.  \cdot(I_{l}+o_{p}(1))W_{n}^{-1}Q_{n}\left(  Q_{n}^{\prime}%
W_{n}^{-1}Q_{n}\right)  ^{-1}(I_{k}+o_{p}(1))V_{n}^{-1/2}-I_{k}\right\Vert \\
&  \leq\left\Vert V_{n}^{-1/2}V_{n}V_{n}^{-1/2}-I_{k}\right\Vert +\left\Vert
V_{n}^{-1/2}V_{n}V_{n}^{-1/2}\right\Vert o_{p}(1)\\
&  \leq o_{p}(1).
\end{align*}
This is (\ref{vvv}). Therefore, (32) is proved.

Finally, (33) follows as in the proof of Theorem 3 (15). \qquad$\blacksquare$

\bigskip

\noindent\textbf{Proof of Theorem 10: } We proceed by verifying the
conditions of Theorem 2.1 of Newey and McFadden (1994) where $E[\log
f(X_{i},\boldsymbol{\theta})]$ and $L_{n}(\boldsymbol{\theta})$ correspond to
their $Q_{0}(\boldsymbol{\theta})$ and $\widehat{Q}_{n}(\boldsymbol{\theta})$.

Their condition (i) holds by Lemma 2.2 of Newey and McFadden (1994) under our
conditions 2 and 3.

Their condition (ii) is our condition 1.

Their conditions (iii) and (iv) hold by our Theorem 5 under our conditions 1, 3, 4, and Assumption 1. \qquad$\blacksquare$

\bigskip

\noindent\textbf{Proof of Theorem 11: } We start by showing that
\[
E\left[  \frac{\partial}{\partial\boldsymbol{\theta}}\log f(X_{i}%
,\boldsymbol{\theta}_{0})\right]  =0,
\]
which holds by Lemma 3.6 of Newey and McFadden (1994) under our conditions
2(a) and 2(b). By Theorem 10 and condition 1,
$\widehat{\boldsymbol{\theta}}$ is in the interior of $\boldsymbol{\Theta}$
with probability approaching one and the first-order condition (FOC) holds:
\[
\frac{1}{n}\sum_{i=1}^{n}\frac{\partial}{\partial\boldsymbol{\theta}}\log
f(X_{i},\boldsymbol{\widehat{\theta}})=0.
\]
Define
\[
R_{n}^{\ast}=V_{n}^{1/2}R_{n}\left(  R_{n}^{\prime}V_{n}R_{n}\right)  ^{-1/2}%
\]
and
\[
\widehat{S}_{n}(\boldsymbol{\theta})=\frac{1}{n}\sum_{i=1}^{n}\frac{\partial
}{\partial\boldsymbol{\theta}}\log f(X_{i},\boldsymbol{\theta}).
\]
By the mean value theorem,
\begin{align}
\left(  R_{n}^{\prime}V_{n}R_{n}\right)  ^{-1/2}R_{n}^{\prime}\sqrt{n}\left(
\widehat{\boldsymbol{\theta}}-\boldsymbol{\theta}\right)   &  =R_{n}%
^{\ast\prime}V_{n}^{-1/2}\sqrt{n}\left(  \widehat{\boldsymbol{\theta}%
}-\boldsymbol{\theta}\right) \nonumber\\
&  =-R_{n}^{\ast\prime}V_{n}^{-1/2}\widehat{H}_{n}\left(  \overline
{\boldsymbol{\theta}}\right)  ^{-1}\sqrt{n}\widehat{S}_{n}\left(
\boldsymbol{\theta}_{0}\right)  \label{mleex}%
\end{align}
where $\overline{\boldsymbol{\theta}}$ is a mean value lies on a line segment
joining $\boldsymbol{\theta}_{0}$ and $\widehat{\boldsymbol{\theta}}$. Let
$\mathcal{N}$ be a neighborhood of $\boldsymbol{\theta}_{0}$. Take $n$ large
enough so that $\widehat{\boldsymbol{\theta}}\in\mathcal{N}$ with probability
approaching one.

We first show
\begin{equation}
\widehat{H}_{n}(\overline{\boldsymbol{\theta}})=H_{n}(\boldsymbol{\theta}%
_{0})(I_{k}+o_{p}(1)). \label{Hcon}%
\end{equation}
Since we can write
\[
\widehat{H}_{n}(\overline{\boldsymbol{\theta}})=H_{n}(\boldsymbol{\theta}%
_{0})\left(  I_{k}+H_{n}(\boldsymbol{\theta}_{0})^{-1}\left(  \widehat{H}%
_{n}(\overline{\boldsymbol{\theta}})-H_{n}(\boldsymbol{\theta}_{0})\right)
\right)  ,
\]
it suffices to show
\[
\left\Vert H_{n}(\boldsymbol{\theta}_{0})^{-1}\left(  \widehat{H}%
_{n}(\overline{\boldsymbol{\theta}})-H_{n}(\boldsymbol{\theta}_{0})\right)
\right\Vert \xrightarrow{p}0.
\]
But by the triangle inequality and Theorem 5,
\begin{align}
&  \left\Vert H_{n}(\boldsymbol{\theta}_{0})^{-1}\left(  \widehat{H}%
_{n}(\overline{\boldsymbol{\theta}})-H_{n}(\boldsymbol{\theta}_{0})\right)
\right\Vert \nonumber\\
&  \leq C^{-1}\left(  \sup_{\boldsymbol{\theta}\in\mathcal{N}}\left\Vert
\widehat{H}_{n}(\boldsymbol{\theta})-H_{n}(\boldsymbol{\theta})\right\Vert
+\left\Vert H_{n}(\overline{\boldsymbol{\theta}})-H_{n}(\boldsymbol{\theta
}_{0})\right\Vert \right)  \xrightarrow{p}0.\nonumber
\end{align}
By Woodbury matrix identity, \eqref{Hcon} implies
\begin{equation}
\widehat{H}_{n}(\overline{\boldsymbol{\theta}})^{-1}=H_{n}(\boldsymbol{\theta
}_{0})^{-1}(I_{k}+o_{p}(1)). \label{H1con}%
\end{equation}

Using \eqref{H1con}, \eqref{mleex} can be written as
\begin{align}
-R_{n}^{\ast\prime}V_{n}^{-1/2}\widehat{H}_{n}\left(  \overline
{\boldsymbol{\theta}}\right)  ^{-1}\sqrt{n}\widehat{S}_{n}\left(
\boldsymbol{\theta}_{0}\right)   &  =-R_{n}^{\ast\prime}V_{n}^{-1/2}%
H_{n}\left(  \boldsymbol{\theta}_{0}\right)  ^{-1}\sqrt{n}\widehat{S}%
_{n}\left(  \boldsymbol{\theta}_{0}\right) \label{qmled2}\\
&  -R_{n}^{\ast\prime}V_{n}^{-1/2}H_{n}\left(  \boldsymbol{\theta}_{0}\right)
^{-1}o_{p}(1)\sqrt{n}\widehat{S}_{n}\left(  \boldsymbol{\theta}_{0}\right)  .
\label{qmled3}%
\end{align}

First take the RHS of \eqref{qmled2}. Since $\text{var}\left(  V_{n}%
^{-1/2}H_{n}\left(  \boldsymbol{\theta}_{0}\right)  ^{-1}\sqrt{n}%
\widehat{S}_{n}\left(  \boldsymbol{\theta}_{0}\right)  \right)  =I_{k}$, by
Theorem 2 under the conditions,
\begin{equation}
-R_{n}^{\ast\prime}V_{n}^{-1/2}H_{n}\left(  \boldsymbol{\theta}_{0}\right)
^{-1}\sqrt{n}\widehat{S}_{n}\left(  \boldsymbol{\theta}_{0}\right)
\xrightarrow{d}N(0,I_{q}). \label{qmleclt}%
\end{equation}
Next, given \eqref{qmleclt}, (\ref{qmled3}) can be bounded by $O_{p}%
(1)o_{p}(1)$. Thus, (34) is proved.

To show (35) it is equivalent to show
\[
\left\Vert V_{n}^{-1/2}\widehat{V}_{n}V_{n}^{-1/2}-I_{k}\right\Vert
\xrightarrow{p}0.
\]
Since \eqref{H1con} holds by replacing $\boldsymbol{\overline{\theta}}$ with
$\boldsymbol{\widehat{\theta}}$,
\begin{equation}
\widehat{H}_{n}(\widehat{\boldsymbol{\theta}})^{-1}=H_{n}\left(
\boldsymbol{\theta}_{0}\right)  ^{-1}\left(  I_{k}+o_{p}(1)\right)  .
\label{H1con1}%
\end{equation}
Since $\lambda_{\min}(\Omega_{n}(\boldsymbol{\theta}))\geq\lambda>0$, with
probability approaching one,
\begin{align}
\widehat{\Omega}_{n}(\widehat{\boldsymbol{\theta}})  &  =\Omega_{n}\left(
\boldsymbol{\theta}_{0}\right)  \left(  I_{k}+\Omega_{n}\left(
\boldsymbol{\theta}_{0}\right)  ^{-1}\left(  \widehat{\Omega}_{n}%
(\widehat{\boldsymbol{\theta}})-\Omega_{n}\left(  \boldsymbol{\theta}%
_{0}\right)  \right)  \right) \nonumber\label{omcon0}\\
&  =\Omega_{n}\left(  \boldsymbol{\theta}_{0}\right)  \left(  I_{k}%
+o_{p}(1)\right)  ,
\end{align}
because
\begin{align}
&  \left\Vert \Omega_{n}(\boldsymbol{\theta}_{0})^{-1/2}\left(
\widehat{\Omega}_{n}(\widehat{\boldsymbol{\theta}})-\Omega_{n}%
(\boldsymbol{\theta}_{0})\right)  \right\Vert \nonumber\\
&  \leq\lambda^{-1}\left(  \sup_{\boldsymbol{\theta}\in\mathcal{N}}\left\Vert
\widehat{\Omega}_{n}(\boldsymbol{\theta})-\Omega_{n}(\boldsymbol{\theta
})\right\Vert +\left\Vert \Omega_{n}(\widehat{\boldsymbol{\theta}}%
)^{-1/2}-\Omega_{n}(\boldsymbol{\theta}_{0})\right\Vert \right) \nonumber\\
&  \leq o_{p}(1).\nonumber
\end{align}
The first inequality holds by the triangle and Schwarz Matrix inequalities.
The second inequality holds by Theorem 6 (23),
$\widehat{\boldsymbol{\theta}}\xrightarrow{p}\boldsymbol{\theta}_{0}$, and
continuity of $\Omega_{n}(\boldsymbol{\theta})$ in $\boldsymbol{\theta}$ under
condition 2(a).

By using \eqref{H1con1} and \eqref{omcon0},
\begin{align}
&  \left\Vert V_{n}^{-1/2}\widehat{V}_{n}V_{n}^{-1/2}-I_{k}\right\Vert
\nonumber\\
&  =\left\Vert V_{n}^{-1/2}\widehat{H}_{n}(\widehat{\boldsymbol{\theta}}%
)^{-1}\widehat{\Omega}_{n}(\widehat{\boldsymbol{\theta}})\widehat{H}%
_{n}(\widehat{\boldsymbol{\theta}})^{-1}V_{n}^{-1/2}-I_{k}\right\Vert
\nonumber\\
&  =\left\Vert V_{n}^{-1/2}H_{n}(\boldsymbol{\theta}_{0})^{-1}\left(  I_{k} +
o_{p}(1)\right)  \Omega_{n}(\boldsymbol{\theta}_{0}) \left(  I_{k}+
o_{p}(1)\right)  H_{n}(\boldsymbol{\theta}_{0})^{-1}\left(  I_{k} +
o_{p}(1)\right)  V_{n}^{-1/2}-I_{k}\right\Vert \nonumber\\
&  \leq\left\Vert V_{n}^{-1/2}V_{n}V_{n}^{-1/2}-I_{k}\right\Vert + \left\Vert
V_{n}^{-1/2}V_{n}V_{n}^{-1/2}\right\Vert o_{p}(1)\nonumber\\
&  \leq o_{p}(1).\nonumber
\end{align}
Thus, (35) is proved.

Finally, (36) follows as in the proof of (15). \qquad
$\blacksquare$

\bigskip

\noindent\textbf{Proof of Theorem 12: } Write $m_{n}%
(\boldsymbol{\theta})=E\overline{m}_{n}(\boldsymbol{\theta})$. Define the
population GMM criterion function as
\[
J_{n}(\boldsymbol{\theta})=n\cdot m_{n}(\boldsymbol{\theta})^{\prime}%
W_{n}^{-1}m_{n}(\boldsymbol{\theta}).
\]
We proceed by verifying the conditions of Theorem 2.1 of Newey and McFadden
(1994). Let $-n^{-1}J_{n}(\boldsymbol{\theta})$ be their $Q_{0}%
(\boldsymbol{\theta})$.

Their condition (i) holds by our conditions 2 and 5.

Their condition (ii) is our condition 1.

By Theorem 5 under Assumption 1 and our conditions 3 and 4,
$m_{n}(\boldsymbol{\theta})$ is continuous in $\boldsymbol{\theta}$ uniformly
over $\boldsymbol{\theta}\in\Theta$ and
\begin{equation}
\label{mulln}\sup_{\boldsymbol{\theta}\in\boldsymbol{\Theta}}\left\Vert
\overline{m}_{n}(\boldsymbol{\theta}) - m_{n}(\boldsymbol{\theta})\right\Vert
\xrightarrow{p}0.
\end{equation}
Since $-n^{-1}J_{n}(\boldsymbol{\theta})$ is continuous and their condition
(iii) holds.

Finally, by $\boldsymbol{\Theta}$ compact, $m_{n}(\boldsymbol{\theta})$ is
bounded on $\boldsymbol{\Theta}$. By the triangle and Schwarz Matrix
inequalities,
\begin{align}
&  \left\Vert \overline{m}_{n}(\boldsymbol{\theta})^{\prime}\widehat{W}%
_{n}^{-1}\overline{m}_{n}(\boldsymbol{\theta})-m_{n}(\boldsymbol{\theta
})^{\prime}W_{n}^{-1}m_{n}(\boldsymbol{\theta})\right\Vert \nonumber\\
&  \leq\left\Vert (\overline{m}_{n}(\boldsymbol{\theta})-m_{n}%
(\boldsymbol{\theta}))^{\prime}\left(  \widehat{W}_{n}^{-1}-W_{n}^{-1}\right)
(\overline{m}_{n}(\boldsymbol{\theta})-m_{n}(\boldsymbol{\theta}))\right\Vert
\nonumber\\
&  + \left\Vert (\overline{m}_{n}(\boldsymbol{\theta})-m_{n}%
(\boldsymbol{\theta}))^{\prime}W_{n}^{-1}(\overline{m}_{n}(\boldsymbol{\theta
})-m_{n}(\boldsymbol{\theta}))\right\Vert + 2\left\Vert (\overline{m}%
_{n}(\boldsymbol{\theta})-m_{n}(\boldsymbol{\theta}))^{\prime}\left(
\widehat{W}_{n}^{-1}-W_{n}^{-1}\right)  m_{n}(\boldsymbol{\theta})\right\Vert
\nonumber\\
&  + \left\Vert m_{n}(\boldsymbol{\theta})^{\prime}\left(  \widehat{W}%
_{n}^{-1}-W_{n}^{-1}\right)  m_{n}(\boldsymbol{\theta})\right\Vert +
2\left\Vert (\overline{m}_{n}(\boldsymbol{\theta})-m_{n}(\boldsymbol{\theta
}))^{\prime}W_{n}^{-1}m_{n}(\boldsymbol{\theta})\right\Vert \nonumber\\
&  \leq\left\Vert \overline{m}_{n}(\boldsymbol{\theta})-m_{n}%
(\boldsymbol{\theta})\right\Vert \left(  \left\Vert \overline{m}%
_{n}(\boldsymbol{\theta})-m_{n}(\boldsymbol{\theta})\right\Vert + 2\left\Vert
m_{n}(\boldsymbol{\theta})\right\Vert \right)  \left(  \left\Vert
\widehat{W}_{n}^{-1}-W_{n}^{-1}\right\Vert + C^{-1}\right) \nonumber\\
&  + \left\Vert m_{n}(\boldsymbol{\theta})\right\Vert ^{2}\left\Vert
\widehat{W}_{n}^{-1}-W_{n}^{-1}\right\Vert .\nonumber
\end{align}
By taking the supremum over $\boldsymbol{\theta}\in\boldsymbol{\Theta}$ on
both sides, their condition (iv) holds by \eqref{mulln} and our condition 6.
\qquad$\blacksquare$

\bigskip

\noindent\textbf{Proof of Theorem 13: } By the conditions 1, 2(a), 3,
and Theorem 12, the sample FOC
\[
2\widehat{Q}_{n}^{\prime}\widehat{W}_{n}^{-1}\overline{m}_{n}%
(\widehat{\boldsymbol{\theta}})=0
\]
is satisfied with probability approaching one. Define
\[
R_{n}^{\ast}=V_{n}^{1/2}R_{n}\left(  R_{n}^{\prime}V_{n}R_{n}\right)
^{-1/2}.
\]
By the mean value theorem, we can write
\begin{align}
\left(  R_{n}^{\prime}V_{n}R_{n}\right)  ^{-1/2}R_{n}^{\prime}\sqrt
{n}(\widehat{\boldsymbol{\theta}}-\boldsymbol{\theta}_{0})  &  =R_{n}%
^{\ast\prime}V_{n}^{-1/2}\sqrt{n}(\widehat{\boldsymbol{\theta}}%
-\boldsymbol{\theta}_{0})\nonumber\\
&  =-R_{n}^{\ast\prime}V_{n}^{-1/2}\left(  \widehat{Q}_{n}^{\prime}%
\widehat{W}_{n}^{-1}\widehat{Q}_{n}(\overline{\boldsymbol{\theta}})\right)
^{-1}\widehat{Q}_{n}^{\prime}\widehat{W}_{n}^{-1}\sqrt{n}\overline{m}%
_{n}(\boldsymbol{\theta}_{0}) \label{focgmm}%
\end{align}
where $\overline{\boldsymbol{\theta}}$ is a mean value lies on a line segment
joining $\boldsymbol{\theta}_{0}$ and $\boldsymbol{\widehat{\theta}}$.

First we show
\begin{equation}
\widehat{Q}_{n}^{\prime}=Q_{n}^{\prime}\left(  I_{l}+o_{p}(1)\right)  .
\label{Qcon}%
\end{equation}
Since $Q_{n}$ is full rank and $\lambda_{\min}(W_{n})\geq C>0$,
\[
\lambda_{\min}\left(  Q_{n}^{\prime}W_{n}Q_{n}\right)  \geq C>0.
\]
We can write
\[
\widehat{Q}_{n}^{\prime}=Q_{n}^{\prime}\left\{  I_{l}+W_{n}^{-1}Q_{n}\left(
Q_{n}^{\prime}W_{n}^{-1}Q_{n}\right)  ^{-1}\left(  \widehat{Q}_{n}%
-Q_{n}(\widehat{\boldsymbol{\theta}})+Q_{n}(\widehat{\boldsymbol{\theta}%
})-Q_{n}\right)  \right\}  .
\]
Let $\mathcal{N}$ be a neighborhood of $\boldsymbol{\theta}_{0}$. Take $n$
large enough so that $\widehat{\boldsymbol{\theta}}\in\mathcal{N}$ with
probability approaching one. By Theorem 5 under the assumptions,
\[
\left\Vert \widehat{Q}_{n}-Q_{n}(\widehat{\boldsymbol{\theta}})\right\Vert
\leq\sup_{\boldsymbol{\theta}\in\mathcal{N}}\left\Vert \widehat{Q}%
_{n}(\boldsymbol{\theta})-Q_{n}(\boldsymbol{\theta})\right\Vert
\xrightarrow{p}0,
\]
and
\[
\left\Vert Q_{n}(\widehat{\boldsymbol{\theta}})-Q_{n}\right\Vert
\xrightarrow{p}0.
\]
Since $Q_{n}=O(1)$,
\[
\left\Vert W_{n}^{-1}Q_{n}\left(  Q_{n}^{\prime}W_{n}^{-1}Q_{n}\right)
^{-1}\left(  \widehat{Q}_{n}-Q_{n}\right)  \right\Vert \leq C^{-2}%
O(1)o_{p}(1).
\]
Thus, \eqref{Qcon} is shown. Using the same argument we also have
\[
\widehat{Q}_{n}^{\prime}(\overline{\boldsymbol{\theta}})=Q_{n}^{\prime}\left(
I_{l}+o_{p}(1)\right)  .
\]

Next, we show
\begin{equation}
\widehat{W}_{n}=W_{n}\left(  I_{l}+o_{p}(1)\right)  . \label{Wcon}%
\end{equation}
We can write
\[
\widehat{W}_{n}=W_{n}\left(  I_{l}+W_{n}^{-1}(\widehat{W}_{n}-W_{n})\right)
.
\]
Recall the definitions (39) and (40) for the weight matrix
$\widehat{W}_{n}(\boldsymbol{\theta})$. By the triangle and Schwarz matrix
inequalities
\begin{align*}
&  \left\Vert W_{n}^{-1}(\widehat{W}_{n}-W_{n})\right\Vert \\
&  \leq\lambda^{-1}\left(  \sup_{\boldsymbol{\theta}\in\mathcal{N}}\left\Vert
\widehat{W}_{n}(\boldsymbol{\theta})-W_{n}(\boldsymbol{\theta})\right\Vert
+\left\Vert W_{n}(\widehat{\boldsymbol{\theta}})-W_{n}\right\Vert \right) \\
&  \leq o_{p}(1)
\end{align*}
since $W_{n}(\boldsymbol{\theta})$ is continuous in $\boldsymbol{\theta}%
\in\mathcal{N}$ (since $m(x,\boldsymbol{\theta})$ is continuously
differentiable) and if
\begin{equation}
\sup_{\boldsymbol{\theta}\in\mathcal{N}}\left\Vert \widehat{W}_{n}%
(\boldsymbol{\theta})-W_{n}(\boldsymbol{\theta})\right\Vert \xrightarrow{p}0.
\label{wulln}%
\end{equation}
Thus, it suffices to show \eqref{wulln}. But this holds by Theorem 5 if $\widehat{W}_{n}(\boldsymbol{\theta})=\frac{1}{n}\sum_{i=1}^{n}%
m(X_{i},\boldsymbol{\theta})m(X_{i},\boldsymbol{\theta})^{\prime}$, by Theorem
4 and 5 if $\widehat{W}_{n}(\boldsymbol{\theta})=\frac{1}%
{n}\sum_{i=1}^{n}m(X_{i},\boldsymbol{\theta})m(X_{i},\boldsymbol{\theta
})^{\prime}-\overline{m}_{n}(\boldsymbol{\theta})\overline{m}_{n}%
(\boldsymbol{\theta})^{\prime}$, and by Theorem 6 if $\widehat{W}%
_{n}(\boldsymbol{\theta})=\widehat{\Omega}_{n}(\boldsymbol{\theta})$. Thus,
\eqref{Wcon} is shown.

By Woodbury matrix identity,
\begin{align}
\widehat{W}_{n}^{-1}  &  =W_{n}^{-1}\left(  I_{l}+o_{p}(1)\right)
,\label{qwqcon}\\
\left(  \widehat{Q}_{n}^{\prime}\widehat{W}_{n}^{-1}\widehat{Q}_{n}%
(\overline{\boldsymbol{\theta}})\right)  ^{-1}  &  =\left(  Q_{n}^{\prime
}W_{n}^{-1}Q_{n}\right)  ^{-1}(I_{k}+o_{p}(1)). \label{W1con}%
\end{align}

By \eqref{Qcon}, \eqref{W1con}, and \eqref{qwqcon}, \eqref{focgmm} can be
written as
\begin{align*}
&  -R_{n}^{\ast\prime}V_{n}^{-1/2}\left(  \widehat{Q}_{n}^{\prime}%
\widehat{W}_{n}^{-1}\widehat{Q}_{n}(\overline{\boldsymbol{\theta}})\right)
^{-1}\widehat{Q}_{n}(\widetilde{\boldsymbol{\theta}})^{\prime}\widehat{W}%
_{n}^{-1}\sqrt{n}\overline{m}_{n}(\boldsymbol{\theta}_{0})\\
&  =-R_{n}^{\ast\prime}V_{n}^{-1/2}\left(  Q_{n}^{\prime}W_{n}^{-1}%
Q_{n}\right)  ^{-1}(I_{k}+o_{p}(1))Q_{n}^{\prime}(I_{l}+o_{p}(1))W_{n}%
^{-1}(I_{l}+o_{p}(1))\sqrt{n}\overline{m}_{n}(\boldsymbol{\theta}_{0})\\
&  =-R_{n}^{\ast\prime}V_{n}^{-1/2}\left(  Q_{n}^{\prime}W_{n}^{-1}%
Q_{n}\right)  ^{-1}Q_{n}^{\prime}W_{n}^{-1}\sqrt{n}\overline{m}_{n}%
(\boldsymbol{\theta}_{0})+U_{n}%
\end{align*}
where $\left\Vert U_{n}\right\Vert =o_{p}(1)$. Since
\[
n\text{var}\left(  -R_{n}^{\ast\prime}V_{n}^{-1/2}\left(  Q_{n}^{\prime}%
W_{n}^{-1}Q_{n}\right)  ^{-1}Q_{n}^{\prime}W_{n}^{-1}\overline{m}%
_{n}(\boldsymbol{\theta}_{0})\right)  =I_{k},
\]
we apply Theorem 2 to find
\[
-R_{n}^{\ast\prime}V_{n}^{-1/2}\left(  Q_{n}^{\prime}W_{n}^{-1}Q_{n}\right)
^{-1}Q_{n}^{\prime}W_{n}^{-1}\sqrt{n}\overline{m}_{n}(\boldsymbol{\theta}%
_{0})\xrightarrow{d}N(0,I_{q}).
\]
Thus, (41) is shown.

To show (42) it is equivalent to show
\[
\left\Vert V_{n}^{-1/2}\widehat{V}_{n}V_{n}^{-1/2}-I_{k}\right\Vert
\xrightarrow{p}0.
\]
Since \eqref{Wcon} holds with $\widehat{W}_{n} = \widehat{\Omega}%
_{n}(\widetilde{\boldsymbol{\theta}})$ and both $\widetilde{
	\boldsymbol{\theta}} $ and $\widehat{ \boldsymbol{\theta}}$ are consistent,
using the same argument with \eqref{Wcon} we have
\begin{equation}
\label{Omcon}\widehat{\Omega}_{n} = \Omega_{n}(I_{l} + o_{p}(1)).
\end{equation}
By using \eqref{Qcon}, \eqref{W1con}, \eqref{qwqcon}, and \eqref{Omcon},
\begin{align}
&  \left\Vert V_{n}^{-1/2}\widehat{V}_{n}V_{n}^{-1/2}-I_{k}\right\Vert
\nonumber\\
&  =\left\Vert V_{n}^{-1/2}(\widehat{Q}_{n}^{\prime}\widehat{W}_{n}%
^{-1}\widehat{Q}_{n})^{-1}\widehat{Q}_{n}^{\prime}\widehat{W}_{n}%
^{-1}\widehat{\Omega}_{n}\widehat{W}_{n}^{-1}\widehat{Q}_{n}(\widehat{Q}%
_{n}^{\prime}\widehat{W}_{n}^{-1}\widehat{Q}_{n})^{-1}V_{n}^{-1/2}%
-I_{k}\right\Vert \nonumber\\
&  \leq\left\Vert V_{n}^{-1/2}V_{n}V_{n}^{-1/2}-I_{k}\right\Vert + \left\Vert
V_{n}^{-1/2}V_{n}V_{n}^{-1/2}\right\Vert o_{p}(1)\nonumber\\
&  \leq o_{p}(1).\nonumber
\end{align}
For the efficient weight matrix case,
\begin{align}
\left\Vert V_{n}^{-1/2}\widehat{V}_{n}V_{n}^{-1/2}-I_{k}\right\Vert  &
=\left\Vert V_{n}^{-1/2}(\widehat{Q}_{n}^{\prime}\widehat{\Omega}_{n}%
^{-1}\widehat{Q}_{n})^{-1}V_{n}^{-1/2}-I_{k}\right\Vert \nonumber\\
&  \leq\left\Vert V_{n}^{-1/2}V_{n}V_{n}^{-1/2}-I_{k}\right\Vert + \left\Vert
V_{n}^{-1/2}V_{n}V_{n}^{-1/2}\right\Vert o_{p}(1)\nonumber\\
&  \leq o_{p}(1).\nonumber
\end{align}
Thus, (42) is proved.

Next, (43) follows as in the proof of (15).

Finally, we show (44). By the mean value theorem, the triangle
inequality, \eqref{Qcon}, and Theorems 2 and 13 (41),
\begin{align}
&  \left\Vert \Omega_{n}^{-1/2}\sqrt{n}\overline{m}_{n}%
(\widehat{\boldsymbol{\theta}})\right\Vert \nonumber\\
&  \leq\left\Vert \Omega_{n}^{-1/2}\sqrt{n}\overline{m}_{n}(\boldsymbol{\theta
}_{0})\right\Vert +\left\Vert \Omega_{n}^{-1/2}Q_{n}\sqrt{n}%
(\widehat{\boldsymbol{\theta}}-\boldsymbol{\theta}_{0})\right\Vert
(1+o_{p}(1))\nonumber\\
&  \leq O_{p}(1)+\left\Vert \sqrt{n}(\widehat{\boldsymbol{\theta}%
}-\boldsymbol{\theta}_{0})^{\prime}V_{n}^{-1/2}V_{n}^{1/2}Q_{n}^{\prime}%
\Omega_{n}^{-1}Q_{n}V_{n}^{1/2}V_{n}^{-1/2}\sqrt{n}%
(\widehat{\boldsymbol{\theta}}-\boldsymbol{\theta}_{0})\right\Vert
^{1/2}(1+o_{p}(1))\nonumber\\
&  \leq O_{p}(1)+\left\Vert V_{n}^{-1/2}\sqrt{n}(\widehat{\boldsymbol{\theta}%
}-\boldsymbol{\theta}_{0})\right\Vert (1+o_{p}(1))\nonumber\\
&  \leq O_{p}(1).\nonumber
\end{align}
Since by \eqref{Omcon} and Woodbury matrix identity,
\begin{align}
\left\Vert n\cdot\overline{m}_{n}(\widehat{\boldsymbol{\theta}})^{\prime
}\widehat{\Omega}_{n}^{-1}\overline{m}_{n}(\widehat{\boldsymbol{\theta}%
})-n\cdot\overline{m}_{n}(\widehat{\boldsymbol{\theta}})^{\prime}\Omega
_{n}^{-1}\overline{m}_{n}(\widehat{\boldsymbol{\theta}})\right\Vert  &
=\left\Vert n\cdot\overline{m}_{n}(\widehat{\boldsymbol{\theta}})^{\prime
}\left(  \widehat{\Omega}_{n}^{-1}-\Omega_{n}^{-1}\right)  \overline{m}%
_{n}(\widehat{\boldsymbol{\theta}})\right\Vert \nonumber\\
&  \leq\left\Vert \Omega_{n}^{-1/2}\sqrt{n}\overline{m}_{n}%
(\widehat{\boldsymbol{\theta}})\right\Vert ^{2}o_{p}(1)\nonumber\\
&  \leq O_{p}(1)o_{p}(1),\nonumber
\end{align}
it suffices to show
\[
n\cdot\overline{m}_{n}(\widehat{\boldsymbol{\theta}})^{\prime}\Omega_{n}%
^{-1}\overline{m}_{n}(\widehat{\boldsymbol{\theta}})\xrightarrow{d}\chi
_{l-k}^{2}.
\]

Using \eqref{focgmm}, \eqref{Qcon}, \eqref{W1con}, and \eqref{qwqcon}, we can
write
\begin{align}
\sqrt{n}(\widehat{\boldsymbol{\theta}}-\boldsymbol{\theta}_{0})  &  =-\left(
\widehat{Q}_{n}^{\prime}\widehat{\Omega}_{n}^{-1}\widehat{Q}_{n}%
(\overline{\boldsymbol{\theta}})\right)  ^{-1}\widehat{Q}_{n}^{\prime
}\widehat{\Omega}_{n}^{-1}\sqrt{n}\overline{m}_{n}(\boldsymbol{\theta}%
_{0})\nonumber\\
&  =-\left(  Q_{n}^{\prime}\Omega_{n}^{-1}Q_{n}\right)  ^{-1}Q_{n}^{\prime
}\Omega_{n}^{-1}\sqrt{n}\overline{m}_{n}(\boldsymbol{\theta}_{0})+o_{p}(1).
\label{gmmex}%
\end{align}
By the mean value theorem and \eqref{gmmex},
\begin{align}
\Omega_{n}^{-1/2}\sqrt{n}\overline{m}_{n}(\widehat{\boldsymbol{\theta}})  &
=\Omega_{n}^{-1/2}\sqrt{n}\overline{m}_{n}(\boldsymbol{\theta}_{0})\nonumber\\
&  -\Omega_{n}^{-1/2}Q_{n}(I_{k}+o_{p}(1))\left(  \left(  Q_{n}^{\prime}%
\Omega_{n}^{-1}Q_{n}\right)  ^{-1}Q_{n}^{\prime}\Omega_{n}^{-1}\sqrt
{n}\overline{m}_{n}(\boldsymbol{\theta}_{0})+o_{p}(1)\right) \nonumber\\
&  =\left(  I_{l}-\Omega_{n}^{-1/2}Q_{n}\left(  Q_{n}^{\prime}\Omega_{n}%
^{-1}Q_{n}\right)  ^{-1}Q_{n}^{\prime}\Omega_{n}^{-1/2}\right)  \Omega
_{n}^{-1/2}\sqrt{n}\overline{m}_{n}(\boldsymbol{\theta}_{0})\label{J1}\\
&  -\Omega_{n}^{-1/2}Q_{n}o_{p}(1)\left(  Q_{n}^{\prime}\Omega_{n}^{-1}%
Q_{n}\right)  ^{-1}Q_{n}^{\prime}\Omega_{n}^{-1}\sqrt{n}\overline{m}%
_{n}(\boldsymbol{\theta}_{0})\label{J2}\\
&  -\Omega_{n}^{-1/2}Q_{n}(I_{k}+o_{p}(1))o_{p}(1). \label{J3}%
\end{align}
Take (\ref{J1}). Since $I_{l}-\Omega_{n}^{-1/2}Q_{n}\left(  Q_{n}^{\prime
}\Omega_{n}^{-1}Q_{n}\right)  ^{-1}Q_{n}^{\prime}\Omega_{n}^{-1/2}$ is
idempotent with rank $l-k$, (\ref{J1}) has the $\chi_{l-k}^{2}$ distribution
asymptotically. For (\ref{J2}),
\begin{align}
&  \left\Vert \Omega_{n}^{-1/2}Q_{n}o_{p}(1)\left(  Q_{n}^{\prime}\Omega
_{n}^{-1}Q_{n}\right)  ^{-1}Q_{n}^{\prime}\Omega_{n}^{-1}\sqrt{n}\overline
{m}_{n}(\boldsymbol{\theta}_{0})\right\Vert \nonumber\\
&  \leq\left\Vert \Omega_{n}^{-1/2}Q_{n}\left(  Q_{n}^{\prime}\Omega_{n}%
^{-1}Q_{n}\right)  ^{-1}Q_{n}^{\prime}\Omega_{n}^{-1/2}\right\Vert
O_{p}(1)o_{p}(1)\nonumber\\
&  \leq o_{p}(1).\nonumber
\end{align}
For (\ref{J3}),
\[
\left\Vert \Omega_{n}^{-1/2}Q_{n}(I_{k}+o_{p}(1))o_{p}(1)\right\Vert
\leq\lambda^{-1/2}O(1)o_{p}(1).
\]
Thus, (44) is shown and the proof is completed. \qquad$\blacksquare$

\end{document}